\RequirePackage{fix-cm}
\documentclass[smallextended]{svjour3}       
\smartqed  
\usepackage[pagewise]{lineno}
\usepackage{array,multirow}
\usepackage{caption}
\usepackage{graphicx}
\usepackage{amsmath}
\usepackage{geometry}
\newcolumntype{P}[1]{>{\centering\arraybackslash}p{#1}}
\newcolumntype{M}[1]{>{\centering\arraybackslash}m{#1}}
\usepackage{tikz}
\usepackage{tikz-qtree}
\usetikzlibrary{trees}
\usepackage{array} 
\usepackage{lscape}

\usepackage{float}
\usepackage{hyperref}
\hypersetup{
	colorlinks=true,linkcolor=blue,anchorcolor=blue,citecolor=blue,filecolor=blue,urlcolor=blue,bookmarksnumbered=true,pdfview=FitB,linktocpage=true,hypertexnames=false,pdftex}

\usepackage[pscoord]{eso-pic}
\newcommand{\placetextbox}[3]{
	\setbox0=\hbox{#3}
	\AddToShipoutPictureFG{ \put(\LenToUnit{#1\paperwidth},\LenToUnit{#2\paperheight}){\vtop{{\null}\makebox[0pt][c]{#3}}}
	}
}
\placetextbox{.5}{0.997}
{\fontsize{6}{6}\selectfont
	This article has been accepted and published in the journal ``Information Processing \& Management". Content may change prior to final publication.  
}
\placetextbox{.5}{0.988}
{\fontsize{6}{6}\selectfont
	DOI: https://doi.org/10.1016/j.ipm.2019.05.009, 0020-0255/ \textcopyright \: 2019 Elsevier Inc. All rights reserved.
}

\placetextbox{.5}{0.035}
{\fontsize{6}{6}\selectfont
	Copyright \textcopyright \: 2019 Elsevier Inc. All rights reserved. This manuscript version is made available under the license http://creativecommons.org/licenses/by-nc-nd/4.0/
}
\begin{document}

\title{Query Expansion Techniques for Information Retrieval: a Survey
}


\author{Hiteshwar Kumar Azad         \and
        Akshay Deepak 
}


\institute{Hiteshwar Kumar Azad \at
              National Institute of Technology Patna \\
              \email{hiteshwar.cse15@nitp.ac.in}           
           \and
            Akshay Deepak \at
              National Institute of Technology Patna \\
              \email{akshayd@nitp.ac.in }
}

\date{Received: date / Accepted: date}

\maketitle

\begin{abstract}
With the ever increasing size of the web, relevant information extraction on the Internet with a query formed by a few keywords has become a big challenge. Query Expansion (QE) plays a crucial role in improving searches on the Internet. Here, the user's initial query is reformulated  by adding additional meaningful terms with similar significance. QE -- as part of  information retrieval (IR)  -- has long attracted researchers' attention. It has become very influential  in the field of personalized social document, question answering, cross-language IR, information filtering and multimedia IR. Research in QE has gained further prominence because of IR dedicated conferences such as TREC (Text Information Retrieval Conference) and CLEF (Conference and Labs of the Evaluation Forum).  This paper surveys QE techniques in IR from 1960 to 2017 with respect to core techniques, data sources used, weighting and ranking methodologies, user participation and applications  -- bringing out similarities and differences. 
\end{abstract}

\section{Introduction}
There is a huge amount of data available on the Internet, and it is growing exponentially. This unconstrained information-growth has not been accompanied by a corresponding technical advancement in the approaches for extracting relevant information \cite{mikroyannidis2007toward}. Often, a web-search does not yield relevant results. There are multiple reasons for this. First, the keywords submitted by the user can be related to multiple topics; as a result, the search results are not focused on the topic of interest. Second, the query can be too short to  capture appropriately what the user is looking for. This can happen just as a matter of habit (e.g., the average size of a web search is 2.4 words \cite{sta,spink2001searching}). Third, the user is often not sure about what he is looking for until he sees the results.   Even if the user knows what he is searching for, he does not know how to formulate an appropriate query (navigational queries are exceptions to this \cite{broder2002taxonomy}). QE plays an important part in fetching relevant results in the above cases. 

Most web queries fall under the following three fundamental categories \cite{broder2002taxonomy,kang2003query} :
\begin{itemize}
	\item \emph{Informational Queries:} Queries that cover a broad topic (e.g., \emph{India} or \emph{journals}) for which there may be thousands of relevant results.
	\item \emph{Navigational Queries:} Queries that are looking for specific website or URL (e.g., \emph{ISRO}).  
	\item \emph{Transactional Queries:} Queries that demonstrate the user's intent to execute a specific activity (e.g., downloading papers or buying books).
\end{itemize}

Currently, user-queries are mostly processed using indexes and ontologies, which work on exact matches and are hidden from the users. This leads to the problem of term mismatch: user queries and search index are not based on the same set of terms. This is also known as the vocabulary problem \cite{furnas1987vocabulary}; it results from a combination of synonymy and polysemy. Synonymy refers to multiple words with common meaning, e.g., ``buy" and ``purchase". Polysemy refers to words with multiple meanings, e.g., ``mouse" (a computer device or an animal). Synonymous and polysemous words are hindrances in retrieving relevant information; they reduce recall and precision rates.  

To address the vocabulary problem, various techniques have been proposed, such as,  relevance feedback, interactive query filtration, corpus dependent knowledge models, corpus independent knowledge models, search result clustering, and word sense disambiguation. Almost all popular techniques expand the initial query by adding new related terms. This can also involve selective retention of terms from the original query. The expanded/reformulated query is then used to retrieve more relevant results. The whole process is called Query expansion (QE). 

Query expansion has a long history in literature. It was first applied in 1960 by Moron and Kuhns \cite{maron1960relevance}  as a technique for literature indexing and searching in a mechanized library system. It was Rocchio \cite{rocchio1971relevance} who brought QE to spotlight through ``relevance feedback'' and its characterization in a vector space model. The idea behind relevance feedback is to incorporate the user's feedback in the retrieval process so as to improve the final result. In particular, the user gives feedback on the retrieved documents in response to the initial query by indicating the relevance of the results. Rocchio's work was further extended and applied in techniques such as collection-based term co-occurrence \cite{jones1971automatic,van1977theoretical}, cluster-based information retrieval \cite{jardine1971use,minker1972evaluation}, comparative analysis of term distribution \cite{porter1982implementing,yu1983generalized,van1986non} and automatic text processing \cite{salton1988term,salton1989automatic,salton1991developments}. 

The above was before the search engine era, where search-retrieval was done on a small amount of data with short queries and satisfactory results were also obtained. In the 1990s, search engines were introduced, and suddenly, huge amounts of data started being published on the web, which has continued to grow at an exponential rate since then. However, users continued to fire short queries for web searches.  While the recall rate suddenly increased,  there was a loss in precision \cite{Salton90improvingretrieval,harman1992relevance}. This called for modernization of QE techniques to deal with Internet-data.  

\begin{figure}[h!]
	\centering 
	\includegraphics[width=13.5cm, height=9cm]{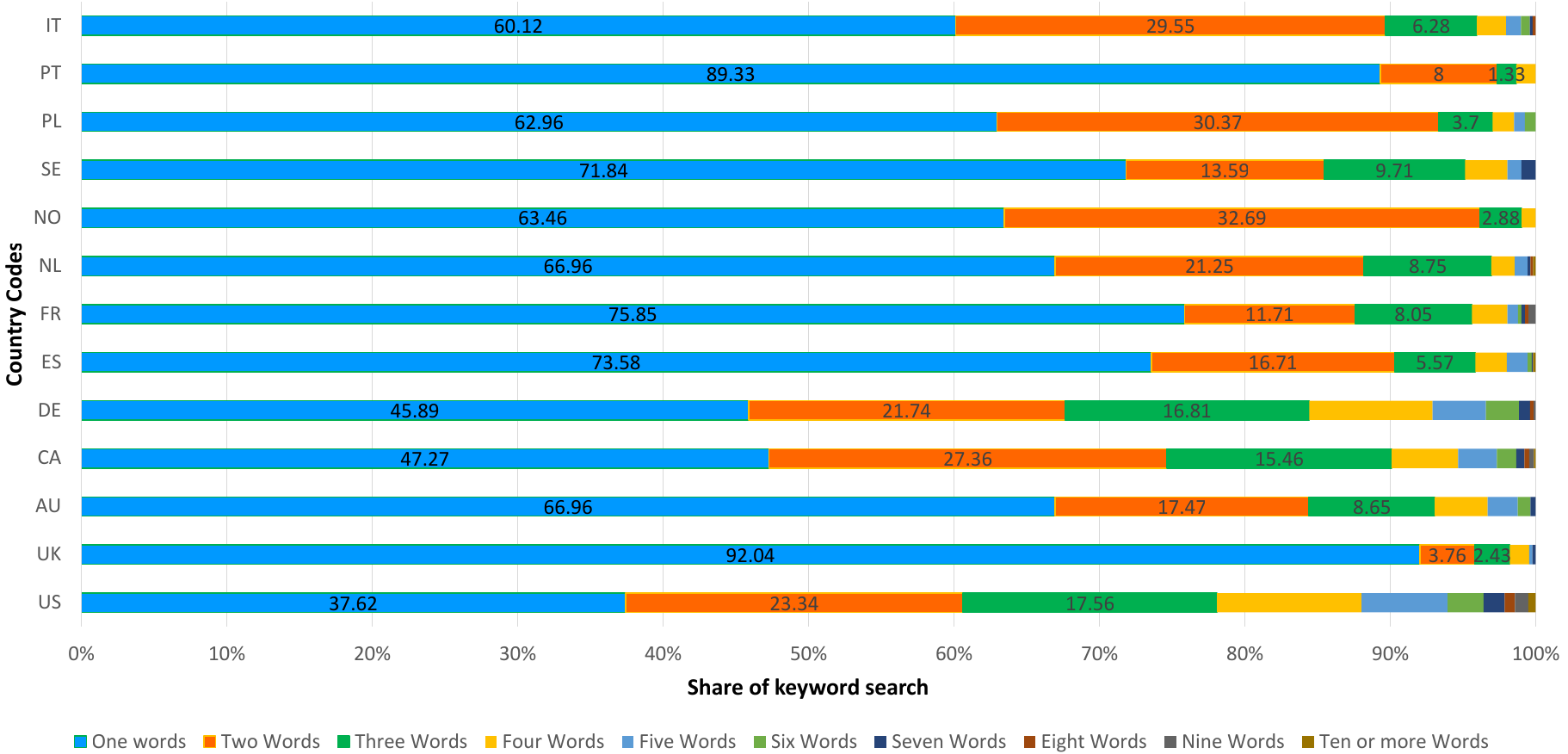}
	\caption{Country-wise size of query searched on Internet constructed using data from \cite{key}}
	\label{fig:b} 
\end{figure}

As per recent reports \cite{sta,key}, the most frequent queries consist of one, two or three words only (see Fig. \ref{fig:b}) -- the same as seventeen years ago as reported by Lau and Horvitz \cite{lau1999patterns}. While the query terms have remained few, the number of web pages have increased exponentially. This has increased the ambiguity -- caused due to the multiple meanings/senses of the query terms (also called vocabulary mismatch problem) -- in finding relevant pages. Hence, the importance of QE techniques has also increased in resolving the vocabulary mismatch problem. 

Recently,  QE has come to spotlight because a lot of researchers are using QE techniques for working on personalized social bookmarking services \cite{ghorab2013personalised,biancalana2013social,bouadjenek2016persador}, Question Answering over Linked Data (QALD)\footnote{http://qald.sebastianwalter.org/} \cite{unger20166th}  and in Text Retrieval Conference (TREC)\footnote{http://trec.nist.gov/}. QE techniques are also used heavily in web, desktop and email searches \cite{pal2015exploring}. Many platforms provide QE facility to end users, which can be turned on or off, e.g., WordNet\footnote{https://wordnet.princeton.edu/}, ConceptNet\footnote{http://conceptnet5.media.mit.edu/}, Lucene\footnote{http://lucene.apache.org/}, Google Enterprise \footnote{https://enterprise.google.com/search/} and MySQL \footnote{https://www.mysql.com/}. 

However, there are also drawbacks of QE techniques, e.g., there is a computational cost associated with the application of QE techniques. In the case of Internet searches, where quick response time is a must, the computational cost associated with the application of QE techniques prohibits their use  in part or entirety  \cite{imran2010selecting}. Another drawback is that sometimes it can fail to establish a relationship between a word in the corpus with those being used in different communities, e.g.,  ``senior citizen" and ``elderly" \cite{gauch1999corpus}. Another issue is that QE may
hurt the retrieval effectiveness for some queries \cite{collins2009reducing,lv2011boosting}. 

\begin{figure}[h!]
	\centering 
	\includegraphics[width=13.5cm, height=10cm]{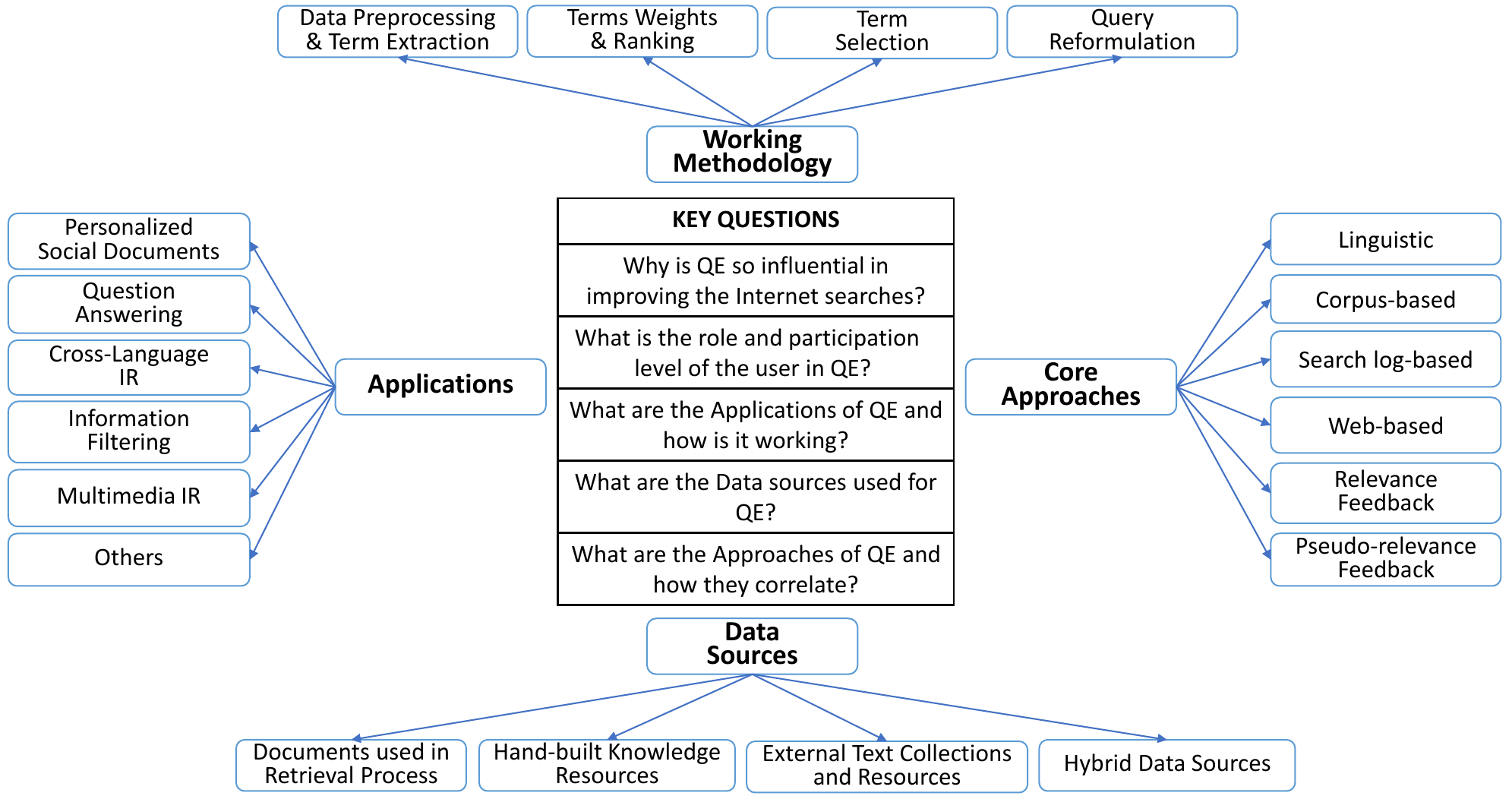}
	\caption{Survey overview}
	\label{fig:a} 
\end{figure}

Few surveys have been done in the past on QE techniques. In 2007, Bhogal et al. \cite{bhogal2007review}  reviewed ontology-based QE techniques, which are domain specific. Such techniques have also been described in book by Manning et al. \cite{Manning:2008:IIR:1394399}.  Carpineto and Romano \cite{carpineto2012survey} (published in the year 2012) reviewed the major QE techniques, Data sources, and features in an IR system. However, their survey covers only automatic query expansion (AQE) techniques and does not include recent research on personalized social documents, term weighting and ranking methods, and categorization of several data sources. After this, we could not find any significant review covering recent progress in QE techniques.   
In contrast, this survey -- in addition to covering recent research in QE techniques -- also covers research on  automatic, manual and interactive QE techniques.  This paper discusses QE techniques from four key aspects: (i) data sources, (ii) applications, (iii) working methodology and (iv) core approaches as summarized in Fig. \ref{fig:a}. 

The rest of the article is organized as follows. Section \ref{sec2} defines QE and describes the working methodology of QE and outlines the main steps. Section \ref{sec3} discuss the importance and application of QE. It also briefly discusses several applications of QE including those in recent literature. Section \ref{sec4} classifies the existing approaches on the basis of properties of various data sources, and comparative analysis of these QE approaches. Finally, Section \ref{sec5} discuss recent trends in literature and concludes the paper.

\section{Query Expansion}
\label{sec2}
Query expansion reformulates the user's original query to enhance the information retrieval effectiveness. Let a user query consist of $n$ terms Q = \{$t_{1}$, $t_2$, ..., $t_i$, $t_{i+1}$, ..., $t_n$\}. The reformulated query can have two components: addition of new terms $T'$=\{$t'_1$, $t'_2$, ..., $t'_m$\} from the data source(s) $D$ and removal of stop words $T''$= \{$t_{i+1}$, $t_{i+2}$, ..., $t_n$ \}.
The reformulated query can be represented as:
\begin{equation}\label{eq:1}
\begin{split}
Q_{exp} & =(Q-T'')\cup T' \\	                                 
& =\{t_{1},t_2, . . .,t_i, t'_1, t'_2,...,t'_m \}
\end{split}
\end{equation}

In the above definition, the key aspect of QE is the set $T'$: set of new meaningful terms added to the user's original query in order to retrieve more relevant documents and reduce ambiguity. 
Karovetz and Croft \cite{krovetz1992lexical} reported that this set $T'$ computed on the basis of term similarity, and without changing the concept, increases recall rate in query results.  Hence, computation of set $T'$ and choice of data sources $D$ are key aspects of research in QE. 

\begin{figure}[h!]
	\centering 
	\includegraphics[width=14cm, height=3.5cm]{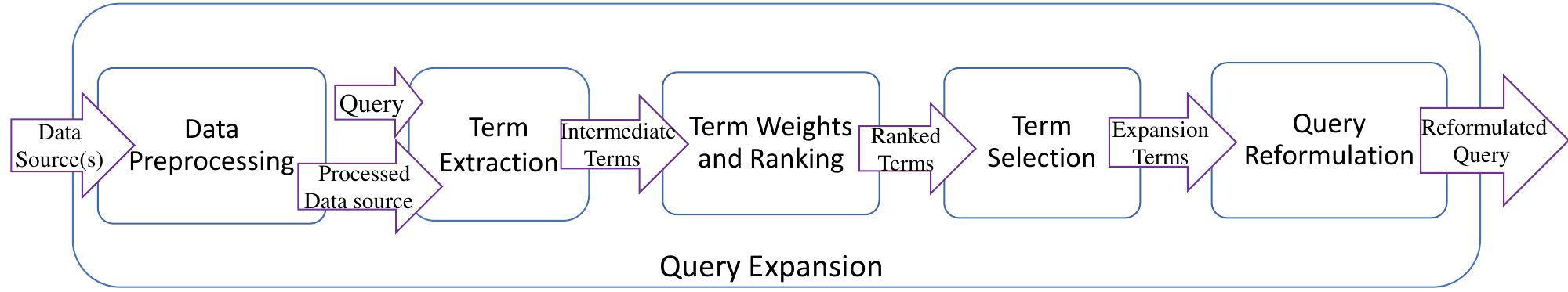}
	\caption{Query expansion process}
	\label{fig:2} 
\end{figure}

In regard to automation and the end-user involvement \cite{efthimiadis1996query}, QE techniques can be classified as follows:
\begin{itemize}
	\item \emph{Manual Query Expansion:} Here, the user manually reformulates the query. 
	\item \emph{Automatic Query Expansion:} Here, the system automatically reformulates the query without any user intervention. Both, the technique to compute set $T'$ and the choice of data sources $D$ is incorporated into the system's intelligence. 
	\item \emph{Interactive Query Expansion:} Here,  query reformulation happens as a result of joint cooperation between the system and the user. It is a human-in-the-loop approach where the system returns search results on an automatically reformulated query, and the users indicate meaningful results among them. Based on the user's preference, the system further reformulates query and retrieves results. The process continues till the user is satisfied with the search results. 	
\end{itemize}
\textbf{Query Expansion Working Methodology:}\\
The process of expanding query consists of four steps: (i) preprocessing of data sources and term extraction, (ii) term weights and ranking, (iii) term selection, and (iv) query reformulation (see Fig. \ref{fig:2}). These steps are discussed next. 

\subsection{Preprocessing of Data Sources and Term Extraction}
Preprocessing of a data source depends upon the data source and the approach being used for QE; it does not depend on the user's query. The primary goal of this step is to extract a set of terms from the data source that meaningfully augment the user's original query.  It consists of the following four sub-steps:
\begin{enumerate}
	\item Text extraction from the data source (extraction of whole texts from the specific data source used for QE)
	\item Tokenization (process of splitting the stream of texts into words)
	\item Stop word removal (removal of frequently used words, e.g., articles, adjective, prepositions, etc.)
	\item Word stemming (process of reducing derived or inflected words to their base word)
\end{enumerate}
After preprocessing the raw data sources, the combined processed data source and the user query are used for term extraction. 

A lot of data sources have been used for QE in literature. All such sources can be classified into four classes: (i) documents used in retrieval process, (ii) hand-built knowledge resources, (iii) external text collections and resources, and (iv) hybrid data sources.
\subsubsection{Documents Used in Retrieval Process} \label{sec1}
At the beginning of the seventies, the addition of similar terms into the initial query started playing a crucial role in QE (e.g.,\cite{jardine1971use,minker1972evaluation,willett1988recent}). Researchers assumed that a set of similar words that frequently appear in documents, belong  to the same subject, thus, similar documents formed a cluster \cite{peat1991limitations}. Two types of clustering have been discussed in document retrieval systems: clustering of terms and clustering of documents \cite{willett1988recent}. A well-known example of term based clustering is Qiu and Frei \cite{qiu1993concept}'s corpus-based expansion technique that uses a similarity thesaurus for expanding the original query. A similarity thesaurus is a collection of documents based on specific domain knowledge, where each term is expressed as a weighted document vector. Another approach proposed by Crouch and Yang \cite{crouch1992experiments} built a statistical corpus thesaurus by clustering the entire document collection using the link clustering algorithm. Some other works that use collection-based data sources for QE are  \cite{jones1971automatic,attar1977local,xu1996query,gauch1999corpus,carpineto2001information,bai2005query}. Carpinto et al. \cite{carpineto2001information} used corpus as data sources from top retrieved documents on the basis of term co-occurrence in the entire document collection. Similarly, Bai et al. \cite{bai2005query} use collection-based data sources as the top-ranked documents and chooses the expansion terms on the basis of term co-occurrence and information flow over the entire corpus. Recently, Zhang et al. \cite{zhang2016learning} used four corpora as data sources (one industry and three academic corpora) and presented a Two-stage Feature Selection framework (TFS) for query expansion known as the Supervised Query Expansion (SQE). The first stage is an Adaptive Expansion Decision (AED), which predicts whether a query is suitable for SQE or not. For unsuitable queries, SQE is skipped with no term features being extracted at all, so that the computation time is reduced. For suitable queries, the second stage conducts Cost Constrained Feature Selection (CCFS), which chooses a subset of effective yet inexpensive features for supervised learning. A drawback of corpus specific QE is that they fail to establish a relationship between a word in the corpus and those which are used in different communities, e.g.,  ``senior citizen" and ``elderly" \cite{gauch1999corpus}.

\subsubsection{Hand-built Knowledge Resources}
The primary goal of hand-built knowledge resources is to extract knowledge from textual hand-built data sources such as dictionaries, thesaurus, ontologies, Wikipedia and LOD cloud. Thesaurus-based QE can be either automatic or hand-built. One of the famous hand-built thesaurus is WordNet \cite{miller1990introduction}. Voorhees \cite{voorhees1994query} utilized WordNet to expand the original query with semantically similar terms called synsets. It was observed that the retrieval effectiveness  improved significantly for unstructured queries, while only marginal improvement was observed for structured queries. Some other articles have also used WordNet to expand the original query, for example,  Smeaton et al. \cite{smeaton1995trec} use synsets of the initial query and assign half weight, Liu et al. \cite{liu2004effective} use word sense, Gong et al. \cite{gong2006multi} use semantic similarity, Zhang et al. \cite{zhang2009concept} use concepts and Pal et al. \cite{pal2014improving} use semantic relations from WordNet. Pal et al. \cite{pal2014improving} propose a new and effective way of using WordNet for QE, where Candidate Expansion Terms (CET) are selected from a set of pseudo-relevant documents and the usefulness of these terms is determined by considering multiple sources of information. The semantic relation between the expanded terms and the query terms is determined using WordNet. Lemos et al. \cite{lemos2014thesaurus} present an automatic query expansion (AQE) approach that uses word relations to increase the chances of finding relevant code. As data sources, it uses a thesaurus containing only software-related word relations and WordNet for expanding the user's query. Similarly, Hsu et al. \cite{hsu2006query} use ConceptNet \cite{liu2004conceptnet} (having higher concept diversity) and WordNet (having higher discrimination ability) as the data sources for expanding the user's query. ConceptNet is a relational semantic network that helps to understand the common sense knowledge of texts written by users. Recently, a number of researchers used ConceptNet as the data source for QE (e.g., \cite{hsu2008combining,kotov2012tapping,bouchoucha2013diversified,anand2015empirical}. Bouchoucha et al. \cite{bouchoucha2013diversified} use ConceptNet for QE and propose a QE technique known as Maximal Marginal Relevance-based Expansion (MMRE). This technique selects expansion terms that are closely related to the initial query but are  different from the previously selected expansion terms. Then, the top N expansion terms having the highest MMRE scores are selected. Recently, Wikipedia and DBpedia are being used widely as data sources for QE (e.g., \cite{li2007improving,arguello2008document,xu2009query,aggarwal2012query,almasri2013wikipedia,anand2015empirical,guisado2016query}). Li et al. \cite{li2007improving} performed an investigation using Wikipedia and retrieved all articles corresponding to the original query as a source of expansion terms for pseudo-relevance feedback. It observed that for a particular query where the general pseudo-relevance feedback fails to improve the query, Wikipedia-based pseudo-relevance feedback improves it significantly. Xu et al. \cite{xu2009query} utilized Wikipedia to categorize the original query into three types: (1) ambiguous queries (queries with terms having more than one potential meaning), (2) entity queries (queries having a specific sense that cover a narrow topic)  and (3) broader queries (queries having neither ambiguous nor specific meaning). They consolidated the expansion terms into the original query and evaluated these techniques using language modeling IR. Almasri et al. \cite{almasri2013wikipedia} use Wikipedia for semantic enrichment of short queries based on in-link and out-link articles.

Augenstein et al. \cite{augenstein2013mapping} use  LOD cloud for keyword mapping and exploits the graph structure within the Linked Data to determine relations between resources that are useful to discover or to express semantic similarity directly. Utilization of data sources as knowledge bases in IR is still an open problem because most of the prior research focuses on the construction of knowledge bases rather than their utilization techniques. Presently, knowledge bases (consisting of entities, their attributes, and their relationships to other entities) are quite popular as data sources for QE. Recently, Xiong and Callan \cite{xiong2015query} use the knowledge base ``freebase" (a large public knowledge base that contains semi-structured information about real-world entities and their facts) for improving  QE. For the selection of the expansion terms,  Xiong and Callan \cite{xiong2015query} developed two methods: (1) utilization of  tf-idf based Pseudo-Relevance Feedback on the linked objects' descriptions, and (2) utilization of Freebase's entity categories, which grant an ontology tree that illustrates entities at several levels of abstraction.

However, Hersh et al. \cite{hersh2000assessing} used a thesaurus relationship for QE in the UMLS Metathesaurus; they reported that nearly all types of QE reduce the recall and precision based on retrieval effectiveness. In their result, not surprisingly, only 38.6\% of the queries with synonym expansion and up to 29.7\% of the queries with hierarchical expansion showed significant improvement in retrieval performance. 
Primarily, there are three limitations in hand-built knowledge resources: they are commonly domain specific, usually do not contain a proper noun and they have to be kept up to date. Experiments with QE using hand-built knowledge resources do not always show  improvements in retrieval effectiveness. It does not improve well-formulated user queries, but significantly improves the retrieval effectiveness of poorly constructed queries.  
\subsubsection{External Text Collections and Resources}
External text collections (such as the WWW, Anchor text, Query logs, External corpus) used in the retrieval process  are the most common and useful data sources for QE. In such cases, QE approaches show overall better performance in comparison to all the discussed data sources. Some data sources under this category need preprocessing procedures for text collection. For example, Kraft and Zien \cite{kraft2004mining}, and, Dang and Croft \cite{dang2010query} use the anchor texts as the data source; they parse hyperlinks to extract data from anchor tags. Further, additional steps need to be carried out such as stop word removal and word stemming. Their experimental results  also suggest that anchor texts can  be used to improve the traditional QE based on query logs. Click through records (URLs, queries) extracted from a search engine (Query logs) is another data source for QE, where users' queries are expanded based on correlation between the query terms and the document terms determined using user logs (e.g., \cite{wen2002query,cui2003query}). Some researchers refer to query logs as user logs since they are derived from  historical records of user queries registered in the query logs of search engines (e.g., \cite{cui2002probabilistic,billerbeck2003query,baeza2004query,yin2009query}). Yin et al. \cite{yin2009query} express the search engine query log as a bipartite graph, where query nodes are connected to the URL nodes by click edges; they reported an improvement of retrieval effectiveness by more than 10\% in average precision. Wang and Zhai \cite{wang2008mining} use web corpus and training data as data sources, and then extract query terms using search logs. Most of the search engines and related surveyed papers using QE are based on query logs. However, for customized search systems for Internet search, enterprise search, personalized search (such as the desktop or email search), or for infrequent queries, query logs are either not available or the user's past queries are not sufficient to describe the information needed. To overcome this limitation, Bhatia et al. \cite{bhatia2011query} proposed a document-centric probabilistic model to generate query suggestions from the corpus that does not depend on query logs and utilizes only the co-occurrence of terms in the corpus. Besides extracting from the user logs, some of the researchers use the sequence of characters comprising the user's query and the corresponding documents from user clicks on URL. This may be useful to remove unwanted content and to find semantically similar terms \cite{beeferman2000agglomerative}.

Today, word embedding techniques are widely used for QE. Recently, Roy et al. \cite{roy2016using} proposed a word embedding framework based on distributed neural language model word2vec. Based on the framework, it extracted similar terms to a query using the K-nearest neighbor approach. The experimental study was done on standard TREC ad-hoc data; it showed considerable improvement over the classic term overlapping-based retrieval approach. It should also be noticed that word2vec based QE methods perform more or less the same with and without any feedback information. Some other works using word embedding techniques are \cite{diaz2016query,kuzi2016query}. Diaz et al. \cite{diaz2016query} presented a QE technique based on locally-trained word embedding (such as word2vec and GloVe) for ad hoc IR. They also used local embeddings that capture the nuances of topic-specific languages and are better than global embeddings. They also suggested that embeddings be learned on topically-constrained corpora, instead of large topically-unconstrained corpora. In a query-specific manner, their experimental results suggest towards adopting local embeddings instead of global embedding because of formers potentially superior representation. Similarly, Kuzi et al. \cite{kuzi2016query} proposed a QE technique based on word embeddings that uses Word2Vec’s Continuous Bag-of-Words (CBOW) approach \cite{mikolov2013efficient}; CBOW represents terms in a vector space based on their co-occurrence in text windows. It also presents a technique for integrating the terms selected using word embeddings with an effective pseudo-relevance feedback method.

Recently, fuzzy logic based expansion techniques have also become popular. Singh et al. \cite{singh2016new,singh2016novel} used a fuzzy logic-based QE technique, and, the top-retrieved documents (obtained using pseudo-relevance feedback) as data sources. Here, each expansion term (obtained from the top retrieved documents) is given a  relevance score using fuzzy rules. The relevance scores of the expanded terms are summed up to infer the high fuzzy weights for selecting expansion terms. 

\subsubsection{Hybrid Data Sources}
Hybrid Data Sources are a combination of two or more data sources (such as the combination of (1) Document used in retrieval process, (2) hand-built knowledge resources, and (3) External text collection and resources). A good number of published works have used hybrid data sources for QE. For example, Collins and Callan \cite{collins2005query} use a combination of query-specific term dependencies from multiple sources such as WordNet,  an external corpus,  and the top retrieved documents as data sources. He and Ounis \cite{he2007combining} use a combination of anchor text, top retrieve documents and corpus as data sources for QE. The main focus is to improve the quality of query term reweighting --  rather than choosing the best terms -- by taking a linear combination of the term frequencies of anchor text, title and body in the retrieved documents. Recently, Pal et al. \cite{pal2013query} used data sources based on term distributions (using Kullback-Leibler Divergence (KLD) and Bose-Einstein statistics (Bo1)) and term association (using Local Context Analysis (LCA) and Relevance-based Language Model (RM3)) methods for QE. The experimental result demonstrated that the combined method gives better result in comparison to each individual method. Other research works based on hybrid resources are \cite{lee2008cluster,Wu:2014:ISR:2556195.2556239,dalton2014entity}. Wu et al. \cite{Wu:2014:ISR:2556195.2556239} use a hybrid data source, which is a combination of three different sources, namely community question answering (CQA) archive, query logs, and web search results. Different types of sources provide different types of signals and reveal the user's intentions from different perspectives. From web search logs they gain an understanding of the wider preference of common web users, from question descriptions they obtain some specific and question-oriented intent, and from the top web search results they further extract some of the popular topics related to the short queries. Dalton et al. \cite{dalton2014entity} propose Entity Query Feature Expansion (EQFE) technique. It uses data sources such as Wikipedia and Freebase to expand the initial query with features from entities and their links to knowledge bases (Wikipedia and Freebase), including structured attributes and text. The main motive for linking entities to knowledge bases is to improve the understanding and representation of text documents and queries. In Anand and Kotov \cite{anand2015empirical}, the document collection and external resources (encyclopedias such as DBpedia and knowledge bases such as ConceptNet) are the data sources for QE. For selecting the expansion terms, term graphs have been constructed  using information theoretic measures based on co-occurrence between each pair of terms
in the vocabulary of the document collection. 
\\\\
\textbf{Comparative Analysis:}
In all the previously discussed data sources, hybrid data sources have been widely used for QE, hence, they can be considered as state-of-art. The main reason behind their widespread acceptance is that they include various features of the user's queries, which cannot be considered by any of the individual data sources. In research involving hybrid data sources, Wikipedia is a popular data source because it is freely available and is the largest encyclopedia on the web, where articles are regularly updated and new articles are added. However, Wikipedia shows  good retrieval effectiveness for short queries only. Data sources belonging to the documents used in the retrieval process have a drawback  that they fail to establish a relationship between a word used in a corpus to words used in the other corpora (e.g.,  ``senior citizen" and ``elderly"). In hand-built knowledge resources, it has been observed that the retrieval effectiveness improved significantly for unstructured queries, while only marginal improvement has been found for structured queries. Mainly, there are three limitations in hand-built knowledge resources: they are usually domain specific, typically do not contain a proper noun and they should be kept up to date. External text collection and resources show overall better performance in comparison to the first two sources discussed earlier. However, some data sources under this category need preprocessing procedure for text collection (e.g., anchor text and query logs). In the case of query logs, it is possible that the query logs are either not available or the user's past queries are not sufficient to describe the information need.

Table \ref{tab:3} summarizes the classification of Data Sources used in QE in literature based on the above discussion.

\begin{table}[!h]
	\centering
	\caption{Summary of Research in Classification of Data Sources used in QE \label{tab:3}}{
		
		\begin{tabular}{ | M{2.2cm} | M{2.2cm} | M{3.3cm} | M{4.5cm} | }
			\hline 
			
			\textbf{Type of Data Sources} & \textbf{Data Sources} & \textbf{Term Extraction Methodology} & \textbf{Publications} \\ \hline 
			\multirow{7}{2.2cm}{\centering {Documents Used in Retrieval Process}} & Clustered terms & Clustering of terms and documents from sets of similar objects & Jardine and Rijsbergen 1971 \cite{jardine1971use}, Minker et al. 1972 \cite{minker1972evaluation}, Willett 1988 \cite{willett1988recent}   \\\cline{2-4}  & Corpus or Collection based data sources & Terms collection from specific domain knowledge & Jones 1971 \cite{jones1971automatic}, Attar and Fraenkel 1977 \cite{attar1977local}, Peat and Willett 1991 \cite{peat1991limitations}, Crouch and Yang 1992 \cite{crouch1992experiments}, Qiu and Frei 1993 \cite{qiu1993concept}, Xu and Croft 1996\cite{xu1996query}, Gauch et al.
			1999 \cite{gauch1999corpus}, Carpineto et al. 2001 \cite{carpineto2001information}, Bai et al. 2005 \cite{bai2005query}  \\ \hline
			\multirow{12}{2.2cm}{\centering {Hand Built Knowledge Resources}} & WordNet \& Thesaurus & Word sense and synset & Miller et al. 1990 \cite{miller1990introduction}, Voorhees
			1994 \cite{voorhees1994query}, Smeaton et al. 1995 \cite{smeaton1995trec}, Liu et al. 2004 \cite{liu2004effective}, Gong et al. 2006 \cite{gong2006multi}, Zhang et al. 2009 \cite{zhang2009concept}, Pal et al. 2014 \cite{pal2014improving}   \\\cline{2-4}  & ConceptNet \& Knowledge bases & Common sense
			knowledge and Freebase & Liu and Singh 2004 \cite{liu2004conceptnet}, Hsu
			et al. 2006 \cite{hsu2006query}, Hsu et al. 2008 \cite{hsu2008combining}, Kotov and Zhai 2012 \cite{kotov2012tapping}, Bouadjenek et al. 2013 \cite{bouadjenek2013sopra}, Anand and Kotov 2015 \cite{anand2015empirical} \\\cline{2-4}  & Wikipedia or DBpedia & Articles, titles \& hyper links & Li et al. 2007 \cite{li2007improving}, Arguello et al. 2008 \cite{arguello2008document}, Xu et al. 2009 \cite{xu2009query}, Aggarwal and Buitelaar 2012 \cite{aggarwal2012query}, ALMasri et al. 2013 \cite{almasri2013wikipedia}, Al-Shboul and Myaeng 2014 \cite{al2014wikipedia}, Anand and Kotov 2015 \cite{anand2015empirical}, Guisado-Gamez et al. 2016 \cite{guisado2016query} 
			\\ \hline
			\multirow{8}{2.2cm}{\centering {External Text Collections and Resources}} & Anchor texts & Adjacent terms in anchor text or text extraction from anchor tags & Kraft and Zien 2004 \cite{kraft2004mining}, Dang and Croft 2010 \cite{dang2010query}   \\\cline{2-4}  & Query logs or User logs & Historical records of user queries registered in the query logs
			of search engine & Wen et al. 2002 \cite{wen2002query}, Cui et al. 2003 \cite{cui2003query}, Billerbeck et al. 2003 \cite{billerbeck2003query}, Baeza-Yates et al. 2004 \cite{baeza2004query}, Yin et al. 2009 \cite{yin2009query}, Hua et al. 2013 \cite{hua2013clickage} \\\cline{2-4}  & External corpus & Nearby terms in word embedding framework & Roy et al. 2016 \cite{roy2016using}, Diaz et al. 2016 \cite{diaz2016query}, Kuzi et al. 2016 \cite{kuzi2016query}, Beeferman et al. 2000 \cite{beeferman2000agglomerative} 
			\\ \hline
			\multirow{0.1}{2.2cm}{\centering {Hybrid Data Sources}}  & Top-ranked documents \& multiple sources & All terms in top retrieved documents  & Collins-Thompson and Callan 2005 \cite{collins2005query}, He and Ounis 2007 \cite{he2007combining}, Lee et al. 2008 \cite{lee2008cluster}, Pal et al. 2013\cite{pal2013query}, Wu et al. 2014 \cite{Wu:2014:ISR:2556195.2556239}, Dalton et al. 2014 \cite{dalton2014entity}, Singh and Sharan 2016 \cite{singh2016new} \\ \hline
			
	\end{tabular}}
\end{table}
\subsection{Weighting and Ranking of Query Expansion Terms}\label{wr}
In this step of QE, weights and ranks are assigned to query expansion terms obtained after data preprocessing (see Fig. \ref{fig:2}). The input to this step is the user's query and texts extracted from the data sources in the first step. Assigned weights denote relevancy of the terms in the expanded query and are further used in ranking retrieved documents based on relevancy. 
There are many techniques for weighting and ranking of query expansion terms. Carpineto and Romano \cite{carpineto2012survey} classify the techniques into four categories on the basis of a relationship between the query terms and the expansion features:
\begin{itemize}
	\item \emph{One-to-One Association:} Correlates each expanded term to at least one query term. 
	\item \emph{One-to-Many Association.} Correlates each expanded term to many query terms.
	\item \emph{Feature Distribution of Top Ranked Documents:} Deals with the top retrieved documents from the initial query and considers the top weighted terms from these documents.
	\item \emph{Query Language Modeling:} Constructs a statistical model for the query and chooses expansion terms having the highest probability.
\end{itemize}
The first two approaches can also be considered as local techniques. These are based on association hypothesis projected by Rijsbergen  \cite{Rijsbergen:1979:IR:539927}: ``If an index term is good at discriminating relevant from non-relevant documents, then any closely associated index term is also likely to be good at this". This hypothesis is primarily motivated by Maron \cite{maron1965mechanized}. Rijsbergen \cite{Rijsbergen:1979:IR:539927} outlines this concept as ``to enlarge the initial request by using additional index terms that have a similar or related meaning to those of the given request". The above approaches have been discussed next.
\subsubsection{One-to-One Association} 
Weighting and ranking the expansion terms based on one-to-one association between the query terms and expansion terms is the most common approach for doing so.  Here, each expansion term is correlated to (at least) one query term (hence the name ``one-to-one"). Weights are assigned to each query term using one of the several techniques described next.

A popular approach to establish a one-to-one association is to use linguistic associations, namely, stemming algorithm. It is used to minimize the inflected word (plural forms, tenses of verbs or derived forms ) from its word-stem. For example, based on Porter's stemming algorithm \cite{porter1980algorithm}, the words ``stems", ``stemmed", ``stemming" and ``stemmer" would be reduced to the root word ``stem." Another typical linguistic approach is the use of thesaurus. One of the most famous thesaurus is WordNet \cite{voorhees1994query}. Using Wordnet, each query term is mapped to its synonyms and a similar set of words -- obtained from WordNet -- in the expanded query. For example, if we consider word ``java" as a noun  in WordNet, there are three synsets with  each having a specific sense: for location (as an island), food (as coffee), and computer science (as a programming language). The same approach has been followed using ConceptNet \cite{hsu2006query} to find the related concepts of a user's queries for query expansion. For example, the word ``file" has a related concept namely ``folder of document", ``record", ``computer", ``drawer", ``smooth rough edge", ``hand tool", etc. Then, each expanded term is assigned a similarity score based on their similarity with the query term. Only terms with high scores are retained in the expanded query.  The natural concept regarding term similarity is that two terms are semantically similar if both terms are in the same document. Similarly, two documents are similar if both are having the same set of terms. There are several approaches to determine term similarity. 

Path length-based measures determine the term similarity between the synsets (senses) -- obtained from WordNet -- based on the path length of the linked synsets. Generally, path length-based measures include two similarity measurement techniques: shortest path similarity \cite{resnik1995using} and  Wu \& Palmer (WP) similarity score \cite{wu1994verbs}. Let the given terms be \emph{$s_1$} and \emph{$s_2$}, and let $len_s$ denote the length of the shortest path between \emph{$s_1$} and \emph{$s_2$} in WordNet. Then, the Shortest path similarity score \cite{resnik1995using} is defined as:
\begin{equation}
Sim_{Path}( s_1, s_2)=\dfrac{1}{len_s} 
\end{equation}
Path length between members of the same synset is considered to be 1; hence, the maximum value of the similarity score can be 1.

\begin{figure}[h!]
	\centering 
	\includegraphics[width=9cm, height=7cm]{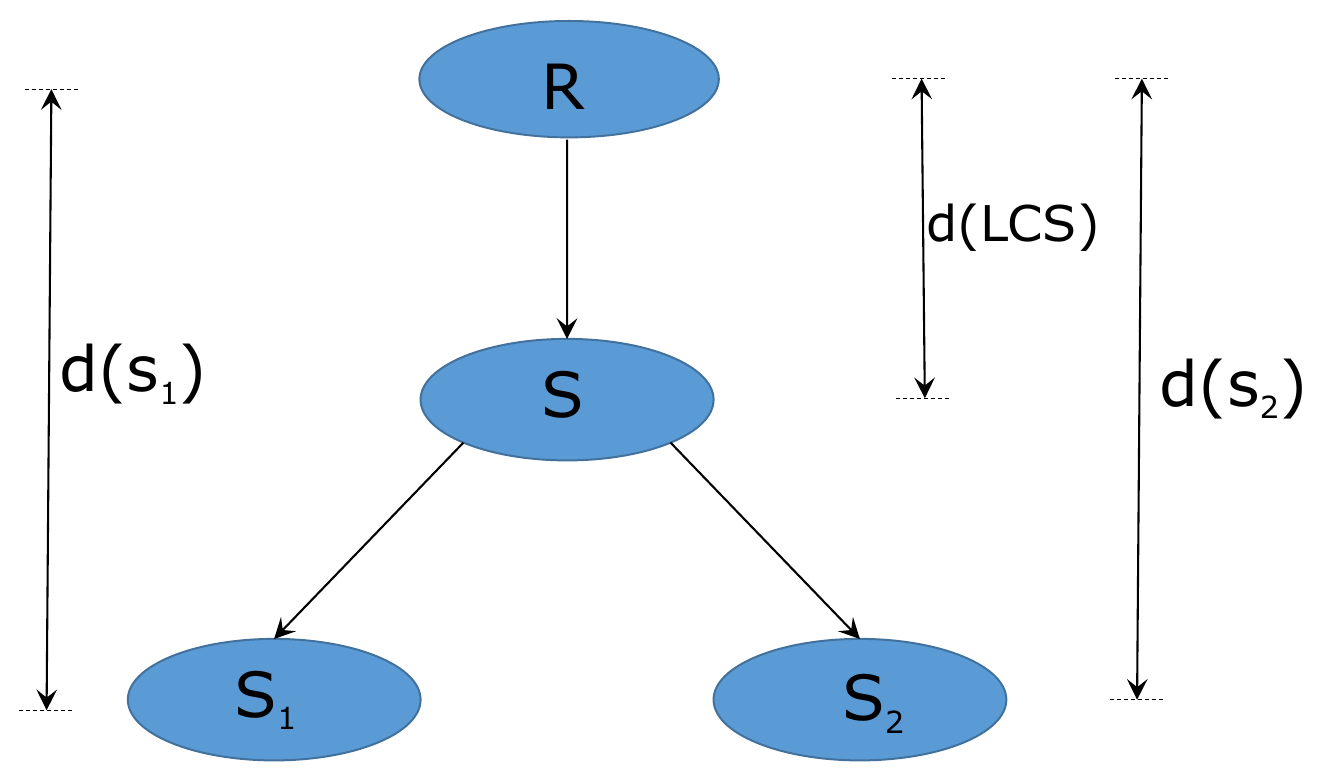}
	\caption{Example of taxonomy hierarchy in WordNet}
	\label{fig:Test2} 
\end{figure}

The Wu \& Palmer (WP) similarity score \cite{wu1994verbs} is defined as
\begin{equation}
Sim_{WP}( s_1, s_2) = \dfrac{2\,.\,d(LCS)}{d(s_1)+d(s_2)}
\end{equation}
where: \\$d(LCS)$ is the depth of Least Common Sub-sumer(LCS)(the closest common ancestor node of two synsets), and\\$d(s_1)$ ($d(s_2)$) is the depth of sense $s_1$ ($s_2$) from the root node R in WordNet (see Figure \ref{fig:Test2}). 

Similarity score of WP similarity varies from 0 to 1 (precisely, 0 $<$ $Sim_{WP}$ $\leq$ 1).

Other approaches like the Jaccard coefficient and Dice coefficient are also used widely for similarity measurement. The Jaccard coefficient \cite{jaccard1912distribution} is described as:
\begin{equation}
Sim_{Jaccard}(s_1, s_2) = \dfrac{df_{s_1\land s_2}}{df_{s_1\lor s_2}}
\end{equation}
where $df_{s_1\land s_2}$ denotes the frequency of documents containing both $s_1$ and $s_2$, and
\\$df_{s_1\lor s_2}$ denotes the frequency of documents containing at least $s_1$ or $s_2$. 

The Dice coefficient \cite{dice1945measures} is described as
\begin{equation}
Sim_{Dice}(s_1, s_2) = \dfrac{2\,.\,df_{s_1\land s_2}}{df_{s_1}+df{s_2}}
\end{equation}
where $df_{s_1}$ and $df_{s_2}$ denote the frequency of documents containing $s_1$ and $s_2$ respectively. 

Term-document matrix is a two dimensional matrix  $M$, whose rows represent the terms and columns represent the documents. Cell $M_{t,d}$ contains value $w_{t,d}$, where $w_{t,d}$ denotes the weight of term $t$ in document $d$. Term-document matrix  is used to compute the similarity score through correlation matrix $C=MM^{T}$, where each cell $c_{s_1, s_2}$ denotes correlation (similarity) score between terms $s_1$ and $s_2$, and is defined as:
\begin{equation}\label{eq:6}
c_{s_1, s_2} = \sum_{d_j}\, w_{s_1, j} \:.\: w_{s_2, j}
\end{equation}
where $w_{s_1, j}$ ($w_{s_2, j}$) is the weight of term $s_1$ ($s_2$) in $j^{th}$ document. 

The cosine similarity measure, denoted $Sim_{cosine}$, is defined as normalization of the above correlation factors:
\begin{equation}
Sim_{cosine}=\frac{c_{s_1,s_2}}{\sqrt{\sum\limits_{d_j}w^2_{s_1,j}\,.\,\sum\limits_{d_j}w^2_{s_2,j}}}
\end{equation}
where:\\ $c_{s_1, s_2}$ denotes correlation (similarity) score between terms $s_1$ and $s_2$, and\\ $w_{s_1, j}$ ($w_{s_2, j}$) is the weight of term $s_1$ ($s_2$) in $j^{th}$ document.

Normalization is done to account for the relative frequency of terms.

It can be seen that using Eq. (\ref{eq:6}) we can create a set of conceptually different term-to-term correlation methods by varying how to select the set of documents and the weighting function. Although calculating co-occurrence of all terms present in the document is easy, it does not consider relative position of terms in a document. For example, two terms that co-occur in the same sentence are more correlated than when they occur in distant parts of a document.

A more exhaustive measurement technique for term co-occurrence that includes term dependency is mutual information \cite{church1990word}:
\begin{equation}\label{eq:8}
I_{s_1, s_2}= log_2 \left[\frac{P(s_1, s_2)}{P(s_1)\,.\, P(s_2)}+ 1\right]
\end{equation}
where: \\$P(s_1, s_2)$ is the combine probability that $s_1$ and $s_2$ co-occur within a particular circumference, and \\$P(s_1)$ and $P(s_2)$ are the respective probabilities of occurrence of terms $s_1$ and $s_2$.

These probabilities can be evaluated by relative frequency count.
Equation \ref{eq:8} is based on mutual information, which is symmetric in nature, i.e., $I(s_1, s_2)= I(s_2, s_1)$. However, in the context of words, order is important, i.e., ``program executing" is different from ``executing program". Hence, it is preferable to consider an asymmetric version of Eq. \ref{eq:8}, where $P(s_1, s_2)$ refers to the probability that $s_1$ exactly follows $s_2$. The mutual information (Eq. \ref{eq:8}) will be: zero if there is no co-occurrence, one if terms $s_1$ and $s_2$ are distinct, and, $log_2 \left(\frac{1}{P(s_1)}+1\right)$ if $s_1$ is completely correlated to $s_2$. 

The drawback of the above formulation is that it can favor infrequent co-occurring terms as compared to frequent distant-occurring terms.

As another option, we can adopt the general description of conditional probability for calculating the stability of association between terms $s_1$ to $s_2$: 
\begin{equation}
P(s_1|s_2)=\dfrac{P(s_1, s_2)}{P(s_2)}
\end{equation}
This well known approach \cite{bai2005query} is identical to the association rule used in data mining problem \cite{agrawal1993mining,vaidya2002privacy}. Association rules have been used widely for identifying the expansion feature correlation with the user query terms \cite{song2007integration,latiri2012towards}. 

Another corpus-based term similarity measure based on information content-based measurement is Resnik similarity \cite{resnik1995using}. Resnik measures the frequent information as information content (IC) of the Least common sub-sumer (LCS) (the closest common ancestor node of two synsets). The value of Resnik similarity would be greater than or equal to zero. The Resnik similarity is formulated as:
\begin{equation}
Sim_{resnik}(s_1, s_2)= -log \,p(LCS(s_1, s_2))
\end{equation} 
where $-log \,p(LCS(s_1, s_2))$ is the information content of the closest common ancestor node of two synsets $s_1$ and $s_2$.

The information content of a synset is defined as the logarithm of the probability of finding the synset in a given corpus.
The negative sign makes the similarity score positive because probabilities are always between [0,1].

Recently, Wikipedia has become popular for expansion of short queries. In Wikipedia, it is possible to have distinct articles with a common title. Every article describes the individual sense of the term, corresponding to the polysemous occurrences of the term in natural language. For example, the term ``apple" has two articles in Wikipedia, one indicating it as a fruit and the other as a company. Almasri et al. \cite{almasri2013wikipedia} use Wikipedia for semantic enhancement of short queries and measure the semantic similarity between two articles $s_1$ and $s_2$ as:
\begin{equation}
Sim_{s_1, s_2}=\dfrac{|I(s_1)\cap I(s_2)|+|O(s_1)\cap O(s_2)|}{|I(s_1)\cup I(s_1)|+|O(s_2)\cup O(s_2)|}
\end{equation} 
where $I(s_1)$ ($I(s_2)$) is the set of articles that point to $s_1$ ($s_2$) as in-links and
\\$O(s_1)$ ($O(s_2)$) is the set of articles that $s_1$ ($s_2$) points to as out-links.

Table \ref{tab:4} summarizes the mathematical form of term similarity score in One-to-One association based on the above discussion.
\begin{table}[!h]
	\centering
	\caption{Summery of One-to-One Association for Term Ranking based on the term similarity score \label{tab:4}}{
		\begin{tabular}{|M{2.2cm}|M{3cm}|M{4.5cm}|}
			\hline
			\textbf{Reference} & \textbf{Approaches} & \textbf{Mathematical form} 
			\\
			\hline 
			
			Jaccard 1912 \cite{jaccard1912distribution} & Jaccard coefficient & $\dfrac{df_{s_1\land s_2}}{df_{s_1\lor s_2}}$ \\
			\hline
			Dice 1945 \cite{dice1945measures} &  Dice coefficient & $\dfrac{2\,.\,df_{s_1\land s_2}}{df_{s_1}+df{s_2}}$\\
			\hline
			Attar and Fraenkel 1977	\cite{attar1977local} & Cosine similarity & $\frac{\sum_{d_j}\, w_{s_1, j} \:.\: w_{s_2,j}} {\sqrt{\sum\limits_{d_j}w^2_{s_1,j}\,.\,\sum\limits_{d_j}w^2_{s_2,j}}}$\\
			\hline
			Church and Hanks 1990 \cite{church1990word} &  Mutual Information & $log_2 \left[\frac{P(s_1, s_2)}{P(s_1)\,.\, P(s_2)}+ 1\right]$  \\
			\hline
			Wu and Palmer 1994 \cite{wu1994verbs} & Wu \& Palmer similarity & 
			$\dfrac{2\,.\,d(LCS)}{d(s_1)+d(s_2)}$
			\\
			\hline
			Resnik 1995 \cite{resnik1995using} &  Resnik similarity & $-log \,p(LCS(s_1, s_2))$\\
			\hline
			ALMasri et al. 2013 \cite{almasri2013wikipedia} & Semantic similarity & $\dfrac{|I(s_1)\cap I(s_2)|+|O(s_1)\cap O(s_2)|}{|I(s_1)\cup I(s_1)|+|O(s_2)\cup O(s_2)|}$ \\
			\hline
			
	\end{tabular}}
\end{table}

\subsubsection{One-to-Many Association} 

In one-to-one association, a candidate term is added to the expanded query if it is correlated to even one term from the original query.  The main issue with one-to-one association is that it may not properly demonstrate the connectivity between the expansion term and the query as a whole.   
For example, consider queries ``data technology" and ``music technology". Here, the word ``technology" is frequently associated with the word ``information". Hence, for query ``data technology", a one-to-one association based expansion of term ``technology"  to ``information technology" may work well because ``information" is strongly correlated to the overall meaning of the query ``data technology". However, the same reasoning does not apply in the case of the query ``music technology". 
Bai et al. \cite{bai2007using} discusses the problem of one-to-one association; it deals with query-specific contexts instead of user-centric ones along with the context around and within the query.

In contrast to one-to-one, in the one-to-many association, a candidate term is added to the expanded query if it is correlated to multiple terms from the original query, hence the name ``one to many". Hsu et al. \cite{hsu2006query,hsu2008combining} use one-to-many association. In these articles, it is compulsory to correlate a new term, extracted from the combination of ConceptNet and WordNet, to a minimum of two original query terms before including the new term into the expanded query. Let $q$ be the original query and let $s_2$ be an expansion term. In one-to-many association, the correlation coefficient of $s_2$ with $q$ is calculated as:
\begin{equation}
\begin{split}
c_{q, s_2} &=\frac{1}{|q|}\sum_{s_1\in q}\, c_{s_1, s_2} \\
& =\frac{1}{|q|}\sum_{s_1\in q}\, \sum_{d_j}\, w_{s_1, j} \:.\: w_{s_2, j}
\end{split}
\label{Eq-One-To-Many}
\end{equation}
where:\\ $c_{s_1, s_2}$ denotes correlation (similarity) score between terms $s_1$ and $s_2$, and\\ $w_{s_1, j}$ ($w_{s_2, j}$) is the weight of the term $s_1$ ($s_2$) in the $j^{th}$ document.

Other works based on one-to-many association are \cite{qiu1993concept,xu1996query,cui2003query,bai2005query,sun2006mining,riezler2008translating,bhatia2011query,gan2015improving,kuzi2016query}. As a special mention, Qiu et al. \cite{qiu1993concept} and Xu et al. \cite{xu1996query} have gained large acceptance in literature because of their  one-to-many association expansion feature and weighting scheme as described in Eq. \ref{Eq-One-To-Many}. 

Qiu et al. \cite{qiu1993concept} use Eq. \ref{Eq-One-To-Many} for finding pairwise correlations between terms in the entire collection of documents. Weight of a term $s$ in document $d_j$, denoted  $w_{s,j}$ (as in Eq. \ref{Eq-One-To-Many}), is computed as the product of term frequency ($tf$) of term $s$ in document  $d_j$ and the inverse term frequency ($itf$) of $d_j$. The $itf$ of  document  $d_j$ is defined as $itf(d_j)=log(\frac{T}{|d_j|})$, where $|d_j|$ is the number of distinct terms in document $d_j$ and $T$ indicates the number of terms in the entire collection. This approach is similar to the inverse document frequency used for document ranking.

Xu et al. \cite{xu1996query} use concepts (a group of contiguous nouns) instead of individual terms while expanding queries. Concepts are chosen based on term co-occurrence with query terms. Concepts are picked from the top retrieved documents, but they are determined on the basis of the top passage (fixed size text window) rather than the whole document. Here, equation Eq. \ref{Eq-One-To-Many} is used for finding the term-concept correlations (instead of term-term correlations), where $w_{s_1, j}$ is the number of co-occurrences of query term $s_1$ in the $j^{th}$ passage and $w_{s_2, j}$ is the frequency of concept $s_2$ in the $j^{th}$ passage. Inverse term frequency of passages and the concepts contained in the passages -- across the entire corpus -- have been considered for calculating the perfect term-concept correlation score. A concept has a correlation factor with every query term. To obtain the correlation factor of the entire query, correlation factors of individual query terms are multiplied. This approach is known as local context analysis \cite{xu1996query}.  

One-to-one association technique tends to be effective only for selecting expansion terms that are loosely correlated to any of the query terms. However, if correlation with the entire query or with multiple query words need to be considered, one-to-many association should be used.  For example, as mentioned before, consider queries ``data technology" and ``music technology".  As discussed before, ``information technology" is not an appropriate expansion for the query ``music technology''. One way to overcome this problem is by adding \emph{context words} to validate term-term associations. For example, in the case of adding ``information" as an expansion term for query ``music technology", an association of ``music" and ``information" should be considered strong only if these terms co-occur together sufficiently high number of times. Here, ``music" is a context word added to evaluate the term-term association of ``information" and ``technology" in the context of the query ``music technology". Such context words can be extracted from a corpus using term co-occurrence \cite{bai2006context,bai2007using,wang2008mining,jian2016simple} or derived from logical significance of a knowledge base \cite{lau2004belief,dalton2014entity,lehmann2015dbpedia,de2016knowledge}.         

Voorhees \cite{voorhees1994query} -- using WordNet data source for QE -- found that the expansion using term co-occurrence techniques is commonly not effective because it doesn't assure a reliable word sense disambiguation. Although, this issue can be resolved by evaluating the correlation between WordNet senses associated with a query term and the senses associated with its neighboring query term. For example, consider query phrase ``incandescent light". In WordNet, the definition of the synset of ``incandescent" contains word ``light". Thus, instead of the phrase ``incandescent light", we can consider the synset of ``incandescent". Liu et al. \cite{liu2004effective} use this approach for Word Sense Disambiguation (WSD). 
For example, consider query ``tropical storm". In WordNet, the sense of the word ``storm" determined through   hyponym of the synset \{violent storm, storm, tempest\} is ``hurricane", whose description has the word ``tropical". As a result, the sense of ``storm" is determined correctly. 

For determining the significance of a phrase, Liu et al. \cite{liu2004effective} evaluate the correlation value of the terms in the phrase. A phrase will be significant if the terms within the phrase are strongly and positively correlated in the collection of documents. The correlation value of the terms in a phrase $\mathcal{P}$ is described as:
\begin{equation}
\label{eq:phraseSig}
C_{s_1, s_2, ..., s_n}=\frac{P(\mathcal{P}) -\prod\limits_{s_i \in \mathcal{P}} P(s_i)}{\prod\limits_{s_i \in \mathcal{P}} P(s_i)}
\end{equation}
where:\\ $s_1, s_2,...,s_n$ are the terms in $\mathcal{P}$,\\ $P(\mathcal{P})$ indicates the probability of a document containing $\mathcal{P}$, \\$P(s_i)$ is the probability of a document containing an individual term $s_i \in \mathcal{P}$,  and \\$\prod\limits_{s_i \in \mathcal{P}} P(s_i)$ indicates the probability of a document having all the terms in $\mathcal{P}$ (assuming that these terms are conditionally independent).

For example, consider a collection of 10 documents where the phrase ``computer science" is present in only one document. Hence, the probability of a document containing the phrase ``computer science" is 0.1. Further, assume that each word in ``computer science" also occurs in only one document -- in which the phrase occurs. Clearly, such a phrase is very significant when part of a query. The same is confirmed  by a high correlation value of 9 computed as per Eq. \ref{eq:phraseSig}. Liu et al. \cite{liu2004effective} suggest the correlation value to be greater than 5 for a phrase to be considered significant. 

Another approach for determining one-to-many association is based on the combination of various relationships between term pairs through a Markov chain framework \cite{collins2005query}. Here, words having the highest probability of relevance in the stationary distribution of the term network are selected for QE. For every individual query term, a term network is built that consists of a pair of correlated terms corresponding to different types of relations (namely synonym, hyponym, co-occurrence). Lv and Zhai \cite{lv2009positional} proposed a positional language model (PLM) that incorporates term proximity evidence in a model-based approach. Term proximity was computed directly based on proximity-based term propagation functions. Song et al. \cite{song2011proximity} proposed Proximity Probabilistic Model (PPM) that uses a position-dependent term count to calculate the number of term occurrences and term counts propagated from neighboring terms. Recently, Jian et al. \cite{jian2016simple} considered term-based information and semantic information as two features of query terms, and presented an efficient ad-hoc IR system using topic modeling. Here, the first topic model is used for extracting the latent semantic information of the query term. Then, term-based information is used as in a typical IR system. This approach is sturdier in relation to data paucity, and it does well on large complicated (belonging to multiple topics) queries. 

To overcome the limitations of term-to-term relationships, one can break the original query as one or more phrases, and then seek for phrases that are similar to it. Phrases usually offer richer context and have less ambiguity in comparison to their individual constituent words. At times, QE even at phrase level may not offer desired clarity. To discuss this further, consider cases of the phrase being compositional or non-compositional. With the compositional phrases, each and every term associated with the phrase can be expanded using similar alternative terms; the final expanded phrase keeps its significance. Cui et al. \cite{cui2003query} analyzes the phrases using n-grams from the user's query logs. The proposed technique filters the phrases that are not present in the documents being searched. Liu et al. \cite{liu2008query} select the most appropriate phrases for QE based on conceptual distance between two phrases (obtained using WordNet). First, phrases similar to the query phrase are selected as candidate phrases. Then, candidate phrases having low conceptual distance with respect to the query phrase are considered in the set of most appropriate phrases. Recently, Al-Shboul and Myaeng \cite{al2014wikipedia} presented a query phrase expansion approach using semantic annotations in Wikipedia pages. It tries to enrich the user query with the phrases that disambiguate the original query word. However, generally, it has been shown that short phrases have a more authentic representation of the information needed, e.g.,  ``artificial intelligence". Further, phrases have a greater inverse document frequency in document collections in the corpus when compared to individual query terms. This is because individual query terms occur more frequently in the document collection than the phrase as a whole. Eguchi \cite{eguchi2005ntcir} acknowledges that retrieval results are improved when pseudo-relevance feedback is also included in QE based on phrases. The combination of pseudo-relevance feedback and phrase expansion is more effective than phrase expansion alone. 

Regarding idiomatic phrases, dealing with them can be troublesome. They are non-compositional in nature and replacing a word with a similar meaning word -- as often done during QE -- can completely change the meaning of the phrase. For example, ``break a leg" is a theatrical slang meaning ``good luck!".  When we replace ``leg" with synonym ``foot", the phrase ``break a foot" gives an  entirely different meaning from the original phrase.

Table \ref{tab:5} summarizes the mathematical form of term-term correlation value in one-to-many association based on the above discussion.
\begin{table}[!h]
	\centering
	\caption{Summary of research work related to one-to-many association QE for term ranking based on term-term correlation values \label{tab:5}}{
		
		\begin{tabular}{|M{4.9cm}|M{3cm}|M{4.9cm}|}
			\hline
			
			\textbf{Publications} & \textbf{Approaches} & \textbf{Mathematical form} 
			\\
			\hline
			
			Qiu and Frei 1993 \cite{qiu1993concept}, Xu and Croft 1996 \cite{xu1996query}, Cui et al. 2003 \cite{cui2003query}, Bai et al. 2005 \cite{bai2005query}, Sun et al. 2006 \cite{sun2006mining}, Riezler et al. 2008 \cite{riezler2008translating}, Bhatia et al. 2011 \cite{bhatia2011query}, Gan and Hong 2015 \cite{gan2015improving}, Kuzi et al. 2016 \cite{kuzi2016query}  & Correlation coefficient & 
			$\begin{array} {lcl} c_{q, s_2} & = & \frac{1}{|q|}\sum_{s_1\in q}\, c_{s_1, s_2} \\
			& = & \frac{1}{|q|}\sum_{s_1\in q} \sum_{d_j}\, w_{s_1, j} .\: w_{s_2, j} \end{array}$ \\
			\hline
			Liu et al. 2004 \cite{liu2004effective} &  Correlation value & $\frac{P(\mathcal{P}) -\prod\limits_{s_i \in \mathcal{P}} P(s_i)}{\prod\limits_{s_i \in \mathcal{P}} P(s_i)}$\\
			\hline			
	\end{tabular}}
\end{table}
\subsubsection{Feature Distribution of Top Ranked Documents} \label{sec:3.2.3}
Approaches discussed in this section are entirely distinct from the approaches described in earlier sections because the QE techniques discussed in this section are not directly associated with the terms (individual or multiple) in the original query. This section uses the top relevant documents for QE in response to the initial query. The idea for using the top retrieved documents as a source of potentially relevant documents for a user's domain of interest comes from Attar and Fraenkel \cite{attar1977local}. The top documents are retrieved in response to the initial query and have more detailed information about the initial query. This detailed information can be used for extracting the most relevant terms for expanding the initial query. Such QE approaches demonstrate collectively better result in comparison to the above approaches. They can be subdivided into two categories:
\begin{itemize}
	\item Query expansion through \emph{Relevance feedback.} Query expansion terms are extracted from the retrieved documents in response to the initial query and the user decides the relevance of the results.
	\item Query expansion through \emph{Pseudo-relevance feedback.} Query expansion terms are extracted from the top-ranked documents in response to the initial query.
	
\end{itemize}

Relevance Feedback (RF) is the most effective QE technique for the modification of  the initial query using the terms extracted from the documents in response to the initial query. The user is asked to assess the relevance of the documents retrieved in response to the initial query. The retrieved documents are shown to the user mostly in some surrogate form, such as title, abstract, keywords, key-phrases or summaries. The user may also have a choice to see the entire documents before making relevant judgment and selecting the relevant documents. After the user indicates relevant documents,  these relevant documents are considered for extracting the terms for the initial QE. The top weighted terms are either added to the initial query automatically or based on manual selection by the user. For example, Java has three synsets with each having a specific sense: island as a geographical place, coffee as a beverage, and as a programming language in computer science. If the query is about Java and the top several retrieved documents are about Java programming, then there may be query drift towards the documents on Java programming. This may not work as desired if the user wants to retrieve documents about Java coffee or Java island. Therefore, if the words added to the original query are unrelated to the query topic, the quality of the retrieval is likely to go down, especially in Web search.

Quite a few term selection techniques have been proposed for QE, which are based on relevance feedback. The common thought behind all these similar techniques is to select terms that will describe the full meaning of the initial query. Rocchio's method \cite{rocchio1971relevance} is one of the first approaches that investigated relevance feedback. This method used an IR system based on the vector space model. The main idea behind this approach is to update the user's initial query vector based on the user's feedback. This method modifies the initial query vector as
\begin{equation}
\overrightarrow{q'}= \alpha .\overrightarrow{q} + \beta .\frac{1}{|RD|}\sum_{\overrightarrow{d_i}\in RD} \overrightarrow{d_i} - \gamma .\frac{1}{|ID|} \sum_{\overrightarrow{d_j}\in ID} \overrightarrow{d_j}
\end{equation}
where:\\
$\overrightarrow{q'}$ is the modified query vector, \\
$\overrightarrow{q}$ is the initial query vector, \\
$\alpha,\beta,\gamma$ manage the comparative importance associated with documents as initial Query  Weight, Relevant Documents (RD) Weight, and irrelevant Documents (ID) Weight respectively, and,\\
$\overrightarrow{d_i},\overrightarrow{d_j}$ are relevant and irrelevant document vectors respectively. \\
In the above paper (i.e.,  \cite{rocchio1971relevance}), only the positive feedback documents and their terms were used to modify and expand the initial query; the weights are typically set as $\alpha=1, \beta=0.75,\gamma=0.15$. Further,  any negative term weights are neglected and set to 0. 

Jones et al. \cite{jones2000probabilistic} present a probabilistic model for calculating document matching score and came up with superior results on TREC Programme collections. Their approach first  retrieves the relevant documents in response to the user's initial query. Then, the documents' Matching Score (MS) is computed as:
\begin{equation}
MS= \sum_{t_i\in q}\,\frac{tf_i \times (k_1+ 1)}{tf_i+NF\times k_1}\times w_i
\end{equation}
where: \\
$t_i$ is an individual term in the user's initial query $q$,\\
$k_1$ is the term frequency normalization factor,\\
$tf_i$ is the term frequency of an individual term $t_i$ in the document,\\ 
$NF$ is the document length normalization factor calculated as $NF= (1-c)+c\times \frac{DL}{AVDL} $ ($c$ is the tuning constant, $DL$ is the document length, and $AVDL$ is average document length), and,\\
$w_i$ is the collection frequency weight of term $t_i$ calculated as $w_i= log \frac{D_N}{n_i}$ ($D_N$ is the total number of documents in the whole collection and $n_i$ is the number of documents containing the term $t_i$).


After the user selects relevant documents in response to the initial query, the system extracts all terms of these documents and ranks them according to  Offer Weight (OW) computed as: 
\begin{equation}
OW=r \times RW
\end{equation}  
where $r$ is the number of relevant documents having the query expansion terms and, $RW$ is the relevance weight, which is computed as:
\begin{equation}
RW= log \frac{(r+0.5)(D_N-n-D_R+r+0.5)}{(D_R-r+0.5)(n-r+0.5)}
\end{equation}
where:\\
$D_N$ is the total number of documents in the collection,\\
$D_R$ is the number of documents selected as relevant by the user, and\\
$n$ is the number of documents containing the term.

After this, either the system asks the user to select relevant terms or adds a fixed number of terms to the user's initial query (automatic query expansion).  

An approach similar to the relevance feedback approach is Pseudo-relevance feedback (or blind feedback, or retrieval feedback). This directly uses the top retrieved documents in response to the user's initial query for composing query expansion terms. Here, the user is not involved in the selection of relevant documents.  Rocchio's method \cite{rocchio1971relevance} can also be applied  in the context of pseudo-relevance feedback, where every individual term extracted from the top retrieved documents is assigned a score by employing a weighting function to the entire collection. The score gathered by every individual term is estimated and the top terms are selected on the basis of the resulting score. The Rocchio's weights can be computed as:
\begin{equation}\label{eq:18}
Score_{Rocchio}(t)= \sum_{d\in R}w(t,d)
\end{equation}
where: \\$w(t,d)$ indicate the weight of term $t$ in pseudo-relevant document $d$ and  \\$R$ is the set of pseudo-relevant documents.

However, a disadvantage of the above approach is that it considers the score of each term in the document collection, and in the process, assigns more importance to the whole collection instead of the user's query. This problem can be resolved by analyzing the term distribution difference between the pseudo-relevant documents and the entire document collection. It is expected that the terms having less information content will have nearly the same distribution in any of the documents from the whole collection. Terms that are closely related to the user's query will have a higher probability of occurrence in the retrieved relevant documents.

Various term ranking functions have been proposed on the basis of term distribution in the pseudo-relevant documents. These functions assign a high score to the terms that differentiate the relevant documents from the irrelevant ones. Some of the important term ranking functions have been described next. 

Robertson et al. \cite{robertson1976relevance} propose a weighting function known as the Binary Independence Model (BIM) that assigns a score to the query terms for term ranking as follows:
\begin{equation}\label{eq:19}
Score_{BIM}(t)= log \frac{p(t|D_R) [1-p(t|D_C)]}{p(t|D_C) [1-p(t|D_R)]}
\end{equation}
where $p(t|D_C)$ and $p(t|D_R)$ signify the probability of occurrence of the term $t$ in the document collection $D_C$ and in a set of pseudo-relevant documents $D_R$ respectively. 

On the same lines, Doszkocs \cite{doszkocs1978aid}  uses a weighting function known as chi-square ($\chi^2$) for scoring the query terms. It is formulated as:

\begin{equation}
Score_{\chi^2}(t)= log \frac{[p(t|D_R)-p(t|D_C)]^2}{p(t|D_C)}
\end{equation}
where $p(t|D_C)$ and $p(t|D_R)$ signify the probability of occurrence of the term $t$ in the document collection $D_C$ and in a set of pseudo-relevant documents $D_R$ respectively.

Robertson \cite{robertson1990term} presents a term selection method based on term weight known as Robertson selection value (RSV). It assigns a weight to a term on the basis of deviation in the term distribution in the top retrieved documents. The term scoring method is formulated as:
\begin{equation}
Score_{RSV}(t)= \sum_{d\in R}w(t,d) . \;[p(t|D_R)-p(t|D_C)]
\end{equation}
where:\\ $w(t,d)$ indicates the weight of the term $t$ in pseudo-relevant document $d$,\\$R$ is the set of pseudo-relevant documents, and,\\ $p(t|D_C)$ and $p(t|D_R)$ signify the probability of occurrence of the term $t$ in the document collection $D_C$ and in a set of pseudo-relevant documents $D_R$ respectively.

On the same lines, Carpineto et al. \cite{carpineto2001information} use the Kullback-Leibler divergence (KLD) for measuring the term distribution difference between pseudo-relevant documents and the entire document collection. Then, the score of the term is computed by adding the score of the terms having higher scores to the KLD score. The score of a term using KLD is computed as:
\begin{equation}
Score_{KLD}(t)=\sum_{t\in V}p(t|D_R).\, log\frac{p(t|D_R)}{p(t|D_C)}
\end{equation}
where $p(t|D_C)$ and $p(t|D_R)$ signify the probability of occurrence of the term $t$ in the document collection $D_C$ and in a set of pseudo-relevant documents $D_R$ respectively.

Using the above term scoring approaches in QE, the experimental studies by Carpineto et al. \cite{carpineto2001information}, Wong et al. \cite{wong2008re}, and Miao et al. \cite{miao2012proximity} showed results with marked improvements.

In 2012, Franzoni and Milani \cite{franzoni2012pming} presented a novel collaborative semantic proximity measurement technique known as PMING distance (further updated in \cite{franzoni2017just}). It is based on the indexing information returned by search engines. It uses the number of occurrences of a term or a set of terms, and counts the number of retrieved results returned by search engines. 

The PMING distance is defined as the weighted combination of Pointwise Mutual Information (PMI) and Normalized Google Distance (NGD). Whereas PMI offers excellent performance in clustering, NGD gives better results in human perception and contexts. Overall, NGD and PMI exhibit good performance in capturing the semantic information for clustering, ranking and extracting meaningful relations among concepts.
In order to understand the PMING distance, we first introduce concept similarity measurement techniques: PMI and NGD.

Pointwise Mutual Information (PMI) \cite{church1990word} is a point-to-point measure of association used in information theory and statistics. Actually, Mutual Information (MI) (Eq. \ref{eq:8}) is a superset of PMI; PMI refers to an individual event, while MI refers to the average of all possible events. PMI is defined in the same manner as MI:
\begin{equation}\label{eq:PMI}
PMI_{s_1, s_2}= log_2 \left[\frac{P(s_1, s_2)}{P(s_1)\,.\, P(s_2)}\right]
\end{equation}

Normalized Google Distance (NGD) \cite{cilibrasi2007google} measures the semantic relation between similar concepts that occur together in a number of documents retrieved by a query on Google or any other search engine. Originally, NGD was developed for Google, but it can be applied to any other search engine. NGD between two terms $s_1$ and $s_2$ is defined as:
\begin{equation}\label{eq:NGD}
NGD_{s_1, s_2}= \frac {max \{log\, f(s_1),log\, f(s_2)\}- log\, f(s_1, s_2)}{log\, N- min \{log\, f(s_1),log\, f(s_2)\}}
\end{equation}
where:\\
$f(s_1)$,  $f(s_2)$, and  $f(s_1, s_2)$ denote the number of results returned by the search engine for query sets   $\{s_1\}$,  $\{s_2\}$ and $\{s_1, s_2\}$ respectively, and,\\
$N$ is the total number of documents indexed by the search engine.

$N$ is usually unknown and varies very frequently. Hence, it can be approximated by a value significantly greater than $max \{ f(s_1), f(s_2)\}$.
Though in human perception NGD may stand well as a proximity measurement technique, in a strict sense it cannot be considered as a metric because it does not satisfy the property of triangular inequality.

PMING distance \cite{franzoni2012pming,franzoni2017just} includes the combination of two semantic similarity measurement techniques: PMI and NGD. PMING distance is defined as a convex linear combination of locally normalized PMI and NGD distances. While combining these two normalized distances, their relatives weights are chosen  based on the context of evaluation using, e.g., Vector Space Model (VSM). For two terms $s_1$ and $s_2$ such that $f(s_1) \geq f(s_2)$, PMING distance between $s_1$ and $s_2$ in context $W$ is given as a function $PMING: W \times W \to \lceil0,1\rceil$ and defined as:
\begin{equation}
PMING_{s_1, s_2}= \rho \,\left[1- \left(log\,\frac{f(s_1, s_2)N}{f(s_1)f(s_2)}\right)\frac{1}{\mu_1}\right] + (1-\rho)\left[\frac{log\, f(s_1)- log\, f(s_1, s_2)}{(log \, N- log \, f(s_2))\mu_2}\right]
\end{equation}
where:\\
$\rho$ is a parameter to balance the weight of components such that $0\leq \rho \leq 1$,\\
$N$ is the total number (if known) or estimated number (if unknown) of documents indexed by the search engine,\\
$\mu_1$ and $\mu_2$ are constants; their values depend on the context of evaluation $W$, and are defined as:

\begin{equation}
\mu_1=\max_{{s_1, s_2}\in W} PMI_{s_1, s_2} \end{equation}
\begin{equation} 
\mu_2=\max_{{s_1, s_2}\in W} NGD_{s_1, s_2} 
\end{equation}

PMING offers the advantages of both PMI and NGD; it outperforms the state-of-the-art proximity measures in modeling human perception, modeling contexts and clustering of semantic associations -- regardless of the search engine/repository.

Recently, Paik et al. \cite{paik2014incremental} presented a scoring function that uses two key properties of a query term:  the number of feedback documents having the query term and the rarity of the query term in the whole document collection. The scoring function is defined as:
\begin{equation}
Score(t,F^q)=log_2(df(t,F^q))\times idf(t,C)
\end{equation}
where:\\
$F^q$ is the set of feedback documents for the query $q$,\\
$df(t,F^q)$ indicates the number of documents in $F^q$ having the term $t$, and,\\
$idf$ stands for inverse document frequency, which is defined as $idf(t,C)=log\frac{N}{df(t,C)}$ ($N$ is the number of documents in the whole collection and $df(t,C)$ corresponds to the document frequency of the term $t$ in the collection $C$).

Every term ranking method has its own motivation, and the outcomes offered by their utilization are also distinct. In the case of  specific queries, it has been observed that the organized sets of expansion terms recommended for each query are mostly unrelated to the original query \cite{carpineto2002improving}. However, various experimental analyses (such as by Harman \cite{harman1992relevance}, Salton and Buckley \cite{salton1997improving}, Carpineto et al. \cite{carpineto2001information}, and Miao et al. \cite{miao2012proximity}) observe that the selection of the ranking approach commonly does not have an enormous significance on the system efficiency; it is just an approach to determine the set of terms for QE.

\subsubsection{Query Language Modeling}\label{ss:QLM}
In this approach for QE, a statistical language model is constructed that assigns a probability distribution over the term-collections. Terms with maximum probability are chosen  for QE. This approach is also known as the model-based approach. The two popular foundation language models are relevance model (based on the probabilities of the terms in the relevant documents) \cite{lavrenko2001relevance,croft2013language} and mixture model \cite{zhai2001model}; both are based on the top retrieved documents for QE. 

In the relevance-based language model, Lavrenko and Croft \cite{lavrenko2001relevance} has caught the attention of researchers with their robust probabilistic approach. Their approach assumes that a query $q_i$ and its top relevant documents set $d$ are sampled randomly (identically and independently) from an unknown relevance model $M_{rel}$. It determines the probability of a term in relevant documents collection on the basis of its co-occurrence with the query terms.  For approximating this relevance model, the probability of term $t$ is computed using the conditional probability of the initial query term $q_i$ ($i\in 1, . . . ,n$). The probability of term $t$ in the relevant documents is computed as:
\begin{equation}\label{eq:27}
p(t|M_{rel})=\sum_{\theta_d\in R} p(\theta_d) \, p(t|\theta_d)\prod_{i=1}^{n}p(q_{i}|\theta_d)
\end{equation} 
In the above Eq. \ref{eq:27}, it is assumed that the term $t$ and the query $q_i$ are mutually independent once they elect a unigram distribution $\theta_d$. Recently, this relevance model has been used widely in QE. This model does not depend upon the distribution difference analysis; hence, it can be said that conceptually this model is very much like Rocchio's method \cite{rocchio1971relevance}. The main difference of this model from the Rocchio's is that the top retrieved documents are assigned a weight such that the lower ranked documents have an insignificant impact on the term probability \cite{lavrenko2006real}. 

In the research area of relevance model, several studies have been published \cite{lv2010positional,bendersky2011parameterized,miao2012proximity}. Lv and Zhai \cite{lv2009positional} performed a correlative analysis on several states of pseudo-relevance feedback and concluded that the relevance model is the most efficient method for the selection of expansion terms. Bendersky et al. \cite{bendersky2011parameterized} use external resources for generating features for weighting different types of query concepts and consider the latent concepts for expanding the initial query. Miao et al. \cite{miao2012proximity} proposed a proximity-based feedback model that is based on the traditional Rocchio's model, known as PRoc. It focuses on the proximity of terms rather than the positional information (unlike position relevance model (PRM)\cite{miao2012proximity}). It calculates the weights of the candidate expansion terms by taking their distance from the query terms into account. Metzler and Croft \cite{metzler2007latent} consider term dependencies during QE; the expansion technique is based on Markov random fields model. This model provides a robust framework that includes both term occurrence and proximity-based features. An example of a Markov random field is estimating the number of times the query terms occur within a window of fixed size in an organized or unorganized way. Lv and Zhai \cite{lv2010positional} present a technique for extracting the expansion terms from the feedback documents known as positional relevance model. Here, the focus is on query topics based on the positions of the query terms in the feedback documents. As another step in improving the research on relevance model, Dalton and Dietz \cite{dalton2013neighborhood} present a neighborhood relevance model that uses relevance feedback approaches for recognizing the specialty of entity linking across the document and query collections. Actually, the primary objectives of entity linking are to map a string in a document to its entity in knowledge base and to recognize the disambiguating context inside the knowledge base. Recently, Dang et al. \cite{dang2016context} proposed a context-dependent relevance model that provides an approach to incorporate the feedback through improvement of document language models. For evaluating document language models, it uses the context information on the relevant or irrelevant documents to obtain the weight counts (using BM25 weights \cite{robertson1994some,robertson2004understanding}) of the individual query terms.

Discussing the mixture model method, Zhai and Lafferty \cite{zhai2001model} consider the top-ranked documents extracted from the document collection that have both relevant and irrelevant information. The proposed method is a mixture productive model that integrates the query topic model $p(t|\theta_Q)$ to the collection language model $p(t|C)$. The collection language model is a suitable model for irrelevant information (content) in top-ranked documents. Following this mixture model, the log-likelihood for top-ranked documents is defined as
\begin{equation}
log \, p(T_R|\theta_Q)=\sum_{D\in T_R}\sum_t c(t,D)\, log(\lambda \,p(t|C)+ (1-\lambda)\, p(t,\theta_Q))
\end{equation}
where:\\
$T_R$ is the set of top-ranked documents,\\
$\theta_Q$ is the estimated query model,\\
$c(t,D)$ is the number of occurrences of term $t$ in document $D$, and,\\
$\lambda$ is a weighting parameter with a value between 0 and 1.

After the evaluation of log-likelihood, Expectation Maximization algorithm \cite{dempster1977maximum} is used to estimate the query topic model so that the likelihood of the top-ranked documents is maximized. However, estimating the query topic model is perhaps more difficult than estimating the document model because queries are generally short -- resulting in the inadequacy of retrieved documents.  Comparatively, this mixture model has a stronger theoretical justification, however, estimating the value of weighting parameter $\lambda$ can be a difficult task.
\\\\
\textbf{Comparative Analysis:}
Among the all weighing and ranking techniques discussed earlier, one-to-many association technique has been used widely. However, weighting and ranking techniques depended upon the different characteristics of the query terms and  the data sources used. One-to-one association technique tends to be effective only for selecting expansion terms that are loosely correlated to any of the query terms. However, this may not accurately demonstrate the relationship between an expansion term and the query as a whole. For example, ``break a leg" is a theatrical slang meaning ``good luck!".  When we replace ``leg" with synonym ``foot", the phrase ``break a foot" gives an  entirely different meaning from the original phrase. For resolving such language ambiguity problem, one-to-many association plays a crucial role. It correlates the entire query or considers  multiple terms from the user's query by assigning correlation score. However, for assigning the weight to the individual terms, one-to-one association play a crucial role. In one-to-one association weighting technique, Jaccard coefficient \cite{jaccard1912distribution} and Cosine similarity \cite{attar1977local} are used widely for assigning the weight to the expansion terms. The weighting techniques discussed under the category of \textit{feature distribution of top-ranked documents} are entirely distinct from the rest because they consider the expansion terms that chosen from the top retrieved documents. However, a disadvantage of the above approach is that it considers the score of each term in the document collection, and in the process, assigns more importance to the whole collection instead of the user's query. Among all the weighting techniques in this category, Rocchio \cite{rocchio1971relevance}, Robertson Selection Value (RSV) \cite{robertson1976relevance} and Kullback-Leibler Divergence (KLD) \cite{carpineto2001information} weighting techniques have been used widely for weighting the expansion terms. Query language modeling is a probabilistic weighting technique that assigns a probability distribution over term-collections. Terms with maximum probability are chosen for QE. Recently, this technique has been used widely in QE. This model does not depend upon the distribution difference analysis; hence, it can be said that conceptually this model is very much like Rocchio's method \cite{rocchio1971relevance}.  

Table \ref{tab:6} summarizes some important term similarity scores in mathematical form for term ranking based on the above discussion.
\begin{table}[!h]
	\centering
	\caption{Summery of Approaches for Term Ranking based on the term similarity score \label{tab:6}}{
		\begin{tabular}{|M{3cm}|M{3cm}|M{6.5cm}|}
			\hline
			
			\textbf{Reference} & \textbf{Approach} & \textbf{Mathematical form} 
			\\
			\hline 
			
			Rocchio 1971 \cite{rocchio1971relevance} & Rocchio's weights & $\sum_{d\in R}w(t,d)$ \\
			\hline
			Robertson and Jones 1976 \cite{robertson1976relevance} &  Dice coefficient & $log \frac{p(t|D_R) [1-p(t|D_C)]}{p(t|D_C) [1-p(t|D_R)]}$\\
			\hline
			Doszkocs 1978 \cite{doszkocs1978aid} & Chi-square ($\chi^2$) & $log \frac{[p(t|D_R)-p(t|D_C)]^2}{p(t|D_C)}$\\
			\hline
			Robertson 1990 \cite{robertson1990term} & Robertson selection value (RSV) & $\sum_{d\in R}w(t,d) . \;[p(t|D_R)-p(t|D_C)]$  \\
			\hline
			Carpineto et al. 2001 \cite{carpineto2001information} & Kullback-Leibler divergence (KLD) & 
			$p(t|D_R).\, log\frac{p(t|D_R)}{p(t|D_C)}$
			\\
			\hline
			Zhai and Lafferty 2001 \cite{zhai2001model} & Log-likelihood & $log \, p(T_R|\theta_Q)=\sum_{D\in T_R}\sum_t c(t,D)\, log(\lambda \,p(t|C)+ (1-\lambda)\, p(t,\theta_Q))$ \\ \hline
			Cilibrasi and Vitanyi 2007 \cite{cilibrasi2007google} & Normalized Google Distance (NGD) & $NGD_{s_1, s_2}= \frac {max \{log\, f(s_1),log\, f(s_2)\}- log\, f(s_1, s_2)}{log\, N- min \{log\, f(s_1),log\, f(s_2)\}}$
			\\
			\hline
			Franzoni and Milani 2012 \cite{franzoni2012pming}, Franzoni 2017 \cite{franzoni2017just} & PMING distance & $ PMING_{s_1, s_2}= \rho \,\left[1- \left(log\,\frac{f(s_1, s_2)N}{f(s_1)f(s_2)}\right)\frac{1}{\mu_1}\right] + (1-\rho)\left[\frac{log\, f(s_1)- log\, f(s_1, s_2)}{(log \, N- log \, f(s_2))\mu_2}\right] $ \\
			\hline
			Paik et al. 2014 \cite{paik2014incremental} &  Scoring function  & $Score(t,F^q)=log_2(df(t,F^q))\times idf(t,C)$\\
			\hline
			
	\end{tabular}}
\end{table}

\subsection{Selection of Query Expansion Terms}
In the previous section \ref{wr}, weighting and ranking of expansion terms were discussed. After this step, the top-ranked terms are selected for QE. The term selection is done on an individual basis; mutual dependence of terms is not considered. This may be debatable; however, some experimental studies (e.g, Lin and Murray \cite{lin2005assessing}) suggest that the independence assumption may be empirically equitable.

It may happen that the chosen QE technique produces a large number of expansion terms, but it might not be realistic to use all of these expansion terms. Usually, only a limited number of expansion terms are selected for QE. This is because the IR effectiveness of a query with a small set of expansion terms is usually better than the query having a large set of expansion terms  \cite{salton1997improving}; this happens due to noise reduction. 

While researchers agree that the addition of selective terms improves the retrieval effectiveness, however, the suggested optimum number can vary from a few terms to a few hundred terms. There are different point of views on the number of selective terms to be added: one-third of the original query terms \cite{robertson1993comparison}, five to ten terms \cite{amati2003probabilistic,chang2006query}, 20-30 terms \cite{harman1992relevance,zhang2016learning}, 30-40 terms \cite{paik2014incremental}, few hundreds of terms \cite{bernardini2008fub,wong2008re}, 350-530 terms for each query \cite{buckley1995automatic}. The source of these terms  can be the top retrieved documents or well known relevant documents. It has been found that the addition of these expansion terms improves retrieval effectiveness by 7\% to 25\% \cite{buckley1995automatic}. On the contrary, some studies show that the number of terms used for QE is a less important factor than the terms selected on the basis of types and quality \cite{sihvonen2004subject}. It has been commonly shown that the effectiveness of QE decreases minutely with the number of non-optimum  expansion terms \cite{carpineto2001information}. Most of the experimental studies observed that the number of expansion terms is of low relevance and it varies from query to query \cite{billerbeck2004questioning,buckley2004reliable,billerbeck2003query,cao2008selecting}. It has been observed that the effectiveness of QE (measured as mean average precision) decreases when we consider less than 20 expansion terms \cite{paik2014incremental,zhang2016learning}. Usually, 20-40 terms are the best choice for QE. Zhai and Lafferty \cite{zhai2001model} assign a probability score to each expansion term and select those with a score higher than a fixed threshold value $p$=0.001.

\begin{table}[!h]
	\centering
	\caption{Summery of Terms selection suggested by several researchers \label{tab:7}}{
		\begin{tabular}{|M{4cm}|M{7.8cm}|}
			
			\hline
			
			\textbf{Number of Terms} & \textbf{Reference}
			\\
			\hline 
			
			One third of the terms & Robertson and Willett 1993 \cite{robertson1993comparison}  \\
			\hline
			5 to 10 terms & Amati et al. 2003 \cite{amati2003probabilistic}, Chang et al. 2006 \cite{chang2006query}  \\
			\hline
			20 to 30 terms & Harman 1992 \cite{harman1992relevance}, Zhang et al. 2016 \cite{zhang2016learning} \\
			\hline
			30 to 40 terms & Paik et al. 2014 \cite{paik2014incremental}  \\
			\hline
			Few hundreds terms & Bernardini and Carpineto 2008 \cite{bernardini2008fub}, Wong et al. 2008 \cite{wong2008re} \\
			\hline
			350 to 530 terms &  Buckley et al. 1995 \cite{buckley1995automatic} \\
			\hline
			
	\end{tabular}}
\end{table}

However, instead of considering an optimum number of expansion terms, it may be better to adopt more informed selection techniques. Focus on the selection of the most relevant terms for QE instead of an optimum number of terms yields better results \cite{carpineto2002improving,cao2008selecting}.

For the selection of the expansion terms on the basis of the ranks assigned to the individual term, various approaches have been proposed that exploit the additional information. Carpineto et al. \cite{carpineto2002improving} proposed a technique that uses multiple term ranking functions and selects the most common terms for each query. A similar approach is utilized by Collins and Callan \cite{collins2007estimation}; however, multiple feedback models are constructed from the same term ranking function. This is done by reconsidering documents from the corpus and by creating alternatives of the initial query. The paper also claims that the proposed technique is effective for eliminating the noise from expansion terms. It aims to expand the query terms, which are related to the various query features. Another approach for selecting expansion terms that depend upon the query ambiguity has been proposed by Chirita et al. \cite{chirita2007personalized}. Here, the number of expansion terms depends on the ambiguity of the initial query in the web or the user log; the ambiguity is determined by the clarity score \cite{cronen2002predicting}. Cao et al. \cite{cao2008selecting} use a classifier to recognize relevant and irrelevant expansion terms. Whether the classifier parameter works well or not for labeling the individual expansion terms, depends on the effectiveness of the retrieval performance and the co-occurrence of the query terms. Their study shows that the top retrieved documents contain as many as 65\% harmful terms. For selecting the best expansion terms, Collins  \cite{collins2009reducing} optimized the retrieved data  with respect to uncertainty sets resulting in an optimization problem.

In spite of the above, it has been shown that the majority of existing works on QE \cite{lavrenko2006real,wu2013incremental} only focus on indexing and document optimization for the selection of expansion terms, and neglect the re-ranking score. However, recently a number of  articles \cite{diaz2015condensed,zhang2016learning} supported the re-ranking with valid proof and obtained good retrieval effectiveness. Wu and Fang \cite{wu2013incremental} proposed impact-sorted indexing technique that utilizes a particular index data structure; the technique improves the scoring methods in IR. Lavrenko and Allan \cite{lavrenko2006real} use the pre-calculated pairwise document similarities to reduce the searching time for the expanded queries. However, supporting re-ranking, Diaz \cite{diaz2015condensed} points out that re-ranking can provide nearly identical performance as the results returned from the second retrieval done using the expanded query. This works specifically for precision-oriented metrics. This has also been verified in experimental results of Zhang et al. \cite{zhang2016learning}, who utilize re-ranking as the default approach for IR. They also suggests to add 20 to 30 expansion terms in the initial query to improve the IR effectively.  

\subsection{Query Reformulation}
This is the last step of QE, where the expanded query is reformulated to achieve better results when used for retrieving relevant documents. The reformulation is done based on the weights assigned to the individual terms of the expanded query; this is known as query reweighting.

A popular query reweighting method was proposed by Salton and Buckley \cite{Salton90improvingretrieval}, which is influenced by Rocchio's method \cite{rocchio1971relevance} for relevance feedback and  its consequential developments. It can be formulated as:
\begin{equation}\label{eqn:26}
w'_{t,q_e} \, = \, (1-\lambda)\,.\, w_{t,q}\, +\, \lambda \,.\, W_t
\end{equation}
where:\\
$w'_{t,q_e}$ is the reweighting of term $t$ of the expanded query $q_e$,\\ $ W_t$ is a weight assigned to the expansion term $t$, and,\\ $\lambda$ is the weighting parameter that balances the comparative contribution of the original query  terms ($q$) and the expansion terms.  

When Rocchio's weights (see Eq. \ref{eq:18}) are used for calculating the weights of the QE terms that are extracted from the pseudo-relevant documents, it can be observed that the expanded query vector measured by Eq. \ref{eqn:26} is relevant to the pseudo-relevant documents. This reduces the term distribution difference between the pseudo-relevant documents and the documents having expansion terms reweighted by Rocchio's weighting scheme. The intention is to assign a low weight to a top-ranked term (in an expanded query) if its relevance score with respect to the whole collection of documents is low. A number of experimental results support this observation for various languages such as Asian \cite{savoy2005comparative}, European  \cite{darwish2014arabic,larkey2007light,amati2003probabilistic}, Hindi \cite{bhattacharya2016usingword,paik2014incremental}, and other languages \cite{zhang2016learning,paik2014incremental,wong2008re,carpineto2001information}. It has been observed that the reweighting system based on inverse term ranks also provides a favorable outcome \cite{hu2006improving,carpineto2002improving}. Another observation is that the document-based weights used for the original unexpanded query and the term distribution difference-based scores used for expansion terms have different units of measurement. Hence, before using them in Eq. \ref{eqn:26}, their values must be normalized. A number of normalization approaches have been discussed in the survey by Wong et al. \cite{wong2008re}; it was observed that the discussed approaches commonly provide similar outcomes. However, Montague and Aslam \cite{montague2001relevance} observe the need for  a better approach that normalizes not only data but also increases equality among normalized terms, which can be more expressive.

In addition to the above discussion, the value of the weighting parameter ($\lambda$) in Eq. \ref{eqn:26} should be adjusted appropriately for improving retrieval effectiveness. A common choice is to grant more  significance  (e.g., multiply by two) to the user's initial query terms in comparison to the expanded query terms. Another way is to use the query reweighting formula without weighting parameter ($\lambda$) as suggested by Amati et al. \cite{amati2003probabilistic}. Another effective approach is to compute the weights, to be assigned to the expansion terms, query-wise . For example, Lv and Zhai \cite{lv2009positional,lv2010positional} use relevance feedback in combination with a learning approach for forecasting the values of weighting parameter ($\lambda$) for each query and every collection of feedback documents.  They also discuss various techniques -- based on, e.g., length, clarity, and entropy -- to measure the correlation of query terms with the entire collection of documents as well as only with the feedback documents.  
However, Eq. \ref{eqn:26} can also be used for extracting expansion terms from hand-built knowledge resources (such as thesaurus, WordNet and ConceptNet). The weighting score may be assigned on the basis of attributes such as the path length, number of co-occurrences, number of connections and relationship types \cite{jones1995interactive}. For example, Voorhees \cite{voorhees1994query} uses expanded query vector with  eleven concept types sub vectors. Each concept type sub vector that comes inside the noun part of WordNet is assigned individual weights. Examples of used concept type sub vectors are ``original query terms" and ``synonyms". Similarly, Hsu et al. \cite{hsu2008combining} use  activation score for weighting of expanded terms. The weight of an expansion term is often decided by its correlation or similarity with the considered query.

When document ranking is based on the language modeling approach (see section \ref{ss:QLM}), the query reweighting step usually favorably expands the original query. In the language modeling framework, the most relevant documents are the ones that decrease Kullback-Leibler divergence (KLD) between the document language model and the query language model. It is formulated as:
\begin{equation}\label{eqn:27}
Sim_{KLD}(Q,D)\propto\sum_{t\in V} p(t|\theta_Q).\, log\frac{p(t|\theta_Q)}{p(t|\theta_D)}
\end{equation}
where:\\
$\theta_Q$ is the query model (usually calculated using the original query terms), and,\\
$\theta_D$ is the document model. 

Document model  $\theta_D$ is calculated based on unknown terms via probability smoothing techniques, such as Jelinek-Mercer interpolation \cite{jelinek1980interpolated,Jelinek1980}:
\begin{equation}\label{eq:C}
p(t|\theta'_D)=(1-\lambda)\,.\, p(t|\theta_D) \,+\, \lambda \,.\, p(t|\theta_{C})
\end{equation}
where:\\
$p(t|\theta'_D)$ is the probability of term $t$ in $\theta'_D$ (documents retrieved using expanded query), and \\
$\theta_{C}$ is the collection model.

Equation \ref{eqn:27} raises the following question: is it possible to build a better query model by obtaining similar terms using their concern-probabilities? Further, will it smoothen the original query model using the equivalent expanded query model (EQM) just as collection model $\theta_{C}$ smoothens the document model based on Eq. \ref{eq:C}? To answer this, several approaches have been proposed to  build an expanded query model that not only considers feedback documents \cite{zhai2001model,lavrenko2001relevance} but also term relations \cite{bai2005query,cao2007extending,gan2015improving} and domain hierarchies \cite{bai2007using}, and can be  heuristic \cite{shah2004evaluating}. Hence, Carpineto and Romano \cite{carpineto2012survey} suggested that instead of considering a particular method, one can come up with a superior expanded query model (calculated using Jelinek-Mercer interpolation \cite{Jelinek1980}) given as:
\begin{equation}
p(t|\theta'_Q)=(1-\lambda)\,.\, p(t|\theta_Q) \,+\, \lambda \,.\, p(t|\theta_{EQM})
\end{equation}
where:\\ 
for each term $t \in \theta'_Q$:
$p(t|\theta'_Q)$ is the probability of term $t$ in the expanded query $\theta'_Q$, \\
$p(t|\theta_Q)$ is the probability of term $t$ in the original query $Q$,\\
$p(t|\theta_{EQM})$ is the probability of term $t$ in the expanded query model $\theta_{EQM}$, and,\\
$\lambda$ is the interpolation coefficient.  

This equation is the probabilistic representation of Eq. \ref{eqn:26} and many articles \cite{xu2009query,kotov2012tapping,carpineto2012survey,gan2015improving,kuzi2016query,zhang2016learning} have used it for probabilistic query reweighting.

Though the query reweighting approach is generally used in QE techniques, it is not mandatory. For example, one can increase the number of similar terms that characterize the original query without using the query reweighting techniques \cite{carpineto2012survey}. Another way can be first to increase the number of similar query terms, and then apply a customized weighting function for ranking the expanded query terms; instead of using the fundamental weighting function used for reweighting the expanded query. This technique was used by Robertson and Walker \cite{robertson2000microsoft} to enhance the Okapi BM25 ranking function \cite{robertson1999okapi}. Some other approaches for query reformulation are utilization of structured query \cite{collins2005query,pound2010expressive,kato2012structured,jamil2015structured}, Boolean query \cite{pane2000improving,liu2004effective,graupmann2005automatic,kim2011automatic}, XML query \cite{kamps2006articulating,chu2006semantic,junedi2012xml} and phrase matching \cite{arguello2008document}.
\section{Importance and Application of Query Expansion}\label{sec3} 
\subsection{Importance of Query Expansion}
One of the major importance of QE is that it enhances the chance to retrieve the relevant information on the Internet, which is not retrieved otherwise using the original query.  Many times the user's original query is not sufficient to retrieve the information user intends or is looking for. In this situation,  QE plays a crucial role in improving Internet searches. For example, if the user's original query is ``\emph{Novel}'', it is not clear what the user wants: the user may be searching for a fictitious narrative book, or the user may be interested in something new or unusual. Here, QE can expand the original query \emph{``Novel"} to \{\emph{``Novel book"}, \emph{``Book"}\}, or to \{\emph{``New"}, \emph{``Novel approach"}\} depending upon the user's interest. The new queries retrieve documents specific to the two types of meaning. This technique has been used hugely for search operations in various commercial domains (e.g., education, hospitality in medical science, economics and experimental research \cite{carpineto2012survey}), where the primary goal is to retrieve all  documents relevant to a particular concern. 

The above fact that the use of QE to retrieve a lot of relevant documents increases recall rate, it adversely affects precision is also well supported experimentally  \cite{harman1996overview,he2009studying,pakhomov2016corpus}. 
The main reason behind the loss in precision is that the relevant documents retrieved in response to the user's initial query may rank lower in the ranking after QE. For improvement of retrieval precision, expanded query can also use Boolean operators (AND,OR) or PARAGRAPH operator \cite{moldovan2000using} to transform the expanded query to Boolean query \cite{pane2000improving,kim2011automatic}, which is eventually submitted for retrieval. For example, let the expanded query (from Eq. \ref{eq:1}) be $T_{exp}$= \{$t_{1}$,$t_2$, . . .,$t_i$, $t'_1$, $t'_2$,...,$t'_m$\}. The expanded Boolean query can be $B_{query}$= \{$t_{1}$ AND $t_2$ AND...AND $t_i$ AND $t'_1$ AND $t'_2$ AND...AND $t'_m$\}, $B_{query}$= \{$t_{1}$ OR $t_2$ OR...OR $t_i$ OR $t'_1$ OR $t'_2$ OR...OR $t'_m$\}, or, a combination of OR and AND operators. A common issue with AND operator is that it improves precision but reduces the recall rate, whereas, OR operator reduces precision but improves the recall rate \cite{turtle1994natural}. Kim et al. \cite{kim2011automatic} propose a novel Boolean query suggestion technique where Boolean queries are produced by exploiting decision trees learned from pseudo-labeled documents. The produced queries are ranked using query quality predictors. The authors compared this technique to contemporary QE techniques and experimentally demonstrated the formers' superiority. XML queries can also be used for improving  precision in IR systems for enhancing the Internet searches (e.g., \cite{kamps2006articulating,chu2006semantic,junedi2012xml}). Improving  precision in IR systems through QE using web pages has been proposed by Cui et al \cite{cui2002probabilistic,cui2003query} and Zhou et al. \cite{zhou2012improving}. Here, QE happens based on a collection of important words in related web pages.  Another set of techniques for the same task expand queries using query concept \cite{qiu1993concept,fonseca2005concept,hsu2008combining,dalton2014entity}. Here, expansion happens based on the similar meaning of query terms.

However, when considering the joint evaluation of the precision and recall rates in QE, several experimental studies agree that the expansion of the user query enhances the average precision of the query results by ten percent or more  (e.g., \cite{salton1997improving,xu2000improving,carpineto2002improving,lee2008cluster,egozi2011concept,rivas2014study,collins2015trec,buttcher2016information}). 
These experimental studies also support the effectiveness of QE techniques in IR systems. Some recent studies have shown that QE can also improve the precision by disambiguating the user query (e.g., \cite{stokoe2003word,bai2005query,zhong2012word,yao2015query}). Table \ref{tab:1} summarizes some prominent works in literature towards improving the precision and recall rates.
\begin{table}[!h]
	\centering
	\caption{Summary of Techniques used for Improving Precision \& Recall rate \label{tab:1}}{
		\begin{tabular}{ |M{5cm} | M{7cm} |}
			
			\hline
			\textbf{Expansion Techniques}    &\textbf{Publications}\\
			\hline
			Boolean query  & Pane and Myers 2000 \cite{pane2000improving}, Moldovan and Mihalcea 2000 \cite{moldovan2000using}, Kim et al. 2011 \cite{kim2011automatic}\\
		\hline
		XML queries	 & Kamps et al. 2006 \cite{kamps2006articulating}, Chu-Carroll et al. 2006 \cite{chu2006semantic}, Junedi et al. 2012 \cite{junedi2012xml}\\
		\hline
		Collection of the top terms within the web pages  & Cui et al. 2002 \cite{cui2002probabilistic}, Cui et al. 2003 \cite{cui2003query}, Zhou et al.
		2012 \cite{zhou2012improving}\\
		\hline
		Query concepts  & Qiu and Frei 1993 \cite{qiu1993concept}, Fonseca et al. 2005 \cite{fonseca2005concept}, Hsu et al. 2008 \cite{hsu2008combining}, Bouchoucha et al. 2013 \cite{bouchoucha2013diversified} \\	
		\hline
		Query Disambiguation & Stokoe et al. 2003 \cite{stokoe2003word}, Bai et al. 2005 \cite{bai2005query}, Zhong and Ng 2012 \cite{zhong2012word}, Yao et al. 2015 \cite{yao2015query}	\\
		\hline	
	\end{tabular}}
\end{table}

\subsection{Application of Query Expansion}
Beyond the key area of IR,  there are other recent applications where QE techniques have proved beneficial. We discuss some of such applications next.

\subsubsection{Personalized Social Documents}
In recent years social tagging systems have achieved popularity by being used in sharing, tagging, commenting, rating, etc., of multi-media contents. Every user wants to find relevant information according to his interests and commitments. This has generated a need of a QE framework that is based on social bookmarking and tagging systems, which enhance the document representation. 

Bender et al. \cite{bender2008exploiting} present a QE framework to exploit the different entities present in social relations (users, documents, tags) and their mutual relationships. It also derives scoring functions for each of the entities and their relations. Biancalana and Micarelli \cite{biancalana2009social} use social tagging and bookmarking in QE for personalized web searches. Their experimental results show effective matching of the user's interests with the search results. Bouadjenek et al. \cite{bouadjenek2011personalized} use a combination of social proximity and semantic similarity for personalized social QE, which is based on similar terms that are mostly used by a given user and his social relatives. Zhou et al. \cite{zhou2012improving} proposed a QE technique that is based on distinctive user profiles in which the expansion terms are extracted from both the annotations and the resources that the user has created and opted (also used by Bouadjenek et al. \cite{bouadjenek2013laicos}). Many other works (e.g., \cite{bouadjenek2013sopra,hahm2014personalized,Mulhem2016}) discuss the QE and social personalized ranking in the context of personalized social documents. Recently, Bouadjenek et al. \cite{bouadjenek2016persador} proposed a technique PerSaDoR (\textbf{Per}sonalized \textbf{S}oci\textbf{a}l \textbf{Do}cument \textbf{R}epresentation) that uses (i) the user's activities in a social tagging system for indexing and modeling, and, (ii) social annotations for QE. A more recent work in personalized IR by Amer et al. \cite{amer2016toward} uses  word embedding for QE. Here, the experimental evaluation was done on the collection of CLEF Social Book Search 2016\footnote{http://social-book-search.humanities.uva.nl}. The main motive of this paper is to address the following questions: (1) ``How to use the word embedding technique for QE in the context of the social collection?" and (2) ``How to use the word embedding technique to personalize QE?". Zhou et al. \cite{zhou2017query} personalized QE using enriched user profiles on the web; the user profiles were created using external corpus ``folksonomy data''. They also proposed a model to enhance the user profiles. This model integrates word embeddings with topic models in two groups of pseudo-relevant documents: user annotations and documents from the external corpus.
\subsubsection{Question Answering}\label{QA}
Question Answering (QA) has become a very influential research area in the field of IR systems. The primary objective of QA is to grant a quick answer in response to the user's query. Here, the focus is to keep the answer concise rather than retrieving all relevant documents. As input, the system accepts questions (instead of a set of terms) in natural language, e.g., ``Which is the first nation in the world to enter Mars orbit in the first attempt?". Recently, search engines have also started using the QA system to provide answers to such types of questions. However, for ranking the answers of such questions, the main challenges in QE is the mismatch problem, which arises due to a mismatch between the expression in question and the text-based answers \cite{lin2001discovery}.

To overcome the mismatch problem and to improve the document retrieval in QA systems, a lot of research has been done. In 2004, Agichtein et al. \cite{agichtein2004learning} presented an important approach for QE using FAQ data. The purposed system automatically learns to transform natural language questions into queries with the goal of maximizing the probability of an information retrieval system returning documents that contain answers to a given question; the same approach was also followed by Soricut and Brill \cite{soricut2006automatic}. Riezler et al. \cite{riezler2007statistical} present a technique to expand the user's original query in QA system using Statistical Machine Translation (SMT) to bridge the lexical gap between questions and answers. SMT attempts to link the linguistic difference between the user's query and system's response. The goal of this system is to learn lexical correlations between words and phrases in questions and answers. Bian et al. \cite{bian2008finding} present a ranking framework to take advantage of user interaction information to retrieve high-quality and relevant content in social media. It has ranked the retrieved documents in QA using  community-based features and, user preference of social media search and web search. Other works \cite{panovich2012tie,liu2015complementary,cavalin2016building,molino2016social} use social network for improving the retrieval performance in QA. Wu et al. \cite{Wu:2014:ISR:2556195.2556239} expand short queries by mining the user intent from three different sources, namely community question answering (CQA) archive, query logs, and web search results. Currently, QE in \textbf{Q}uestion \textbf{A}nswering over \textbf{L}inked open \textbf{D}ata (QALD) has gained much attention in the area of natural language processing. Shekarpour et al. \cite{shekarpour2013keyword} has proposed an approach for expansion of the original query on linked data using linguistic and semantic features, where linguistic features are extracted from WordNet and semantic features are extracted from Linked Open Data (LOD)\footnote{http://lod-cloud.net/} cloud. The evaluation was carried out on a training dataset extracted from the QALD\footnote{http://qald.sebastianwalter.org/} question answering benchmark dataset. The experimental results show a considerable improvement in precision and recall rates over the baseline approaches. 

\subsubsection{Cross-Language Information Retrieval (CLIR)}
It is a part of IR that retrieves the information present in a language(s) different from the user's query-language. For example, a user can query in Hindi, but the retrieved relevant information can be in English. Over the past few years, CLIR has received significant attention due to the popularity of CLEF\footnote{http://www.clef-initiative.eu/} and TREC, which are held annually for promoting the research in the area of IR.

Traditionally, there are three main approaches to CLIR: query translation with machine translation techniques \cite{radwan1994vers}, parallel or comparable corpora-based techniques \cite{sheridan1996experiments} and machine-readable bilingual dictionaries \cite{ballesteros1996dictionary}. Research challenges with the traditional CLIR are untranslatable query terms, phrase translation, inflected term and uncertainty in language translation between the source and target languages \cite{pirkola2001dictionary}. To overcome this translation error, a popular approach is to use QE \cite{ballesteros1997phrasal,nie1999cross}. It gives a better output -- even in the case of no translation error -- due to the use of statistical semantic similarity among the terms \cite{adriani1999term,kraaij2003embedding}. To counter the errors in the automated machine translation in the case of cross-language queries, Gaillard et al. \cite{gaillard2010query} use linguistic resources for QE. QE can be applied at various points in the translation process: before or after translation, or both. It has been shown that the application at prior translation gives better result in comparison to the application at post-translation, however, the application at either step gives superior results in comparison to not using QE \cite{ballesteros1998resolving,ballesteros2002cross,mcnamee2002comparing,levow2005dictionary}. For improving the QE in CLIR, Cao et al. \cite{cao2007extending} combines the dictionary translation and the co-occurrence term-relations into Markov Chain (MC) models, which is defined as a directed graph where query translation is formulated as a random walk in the MC models. Recently, Zhou et al.  \cite{zhou2015query,zhou2016study} used QE techniques to personalize CLIR based on user's historical usage information. Experimental results show that personalized approaches work better than non-personalized approaches in CLIR.  Bhattacharya et al.  \cite{bhattacharya2016usingword} present a technique to translate user queries from Hindi to English CLIR using word embedding. The proposed word embedding based approach captures contextual clues for a particular word in the source language and gives those words as translations that occur in a similar context in the target language.

\subsubsection{Information Filtering} 
Information filtering (IF) is a method to eliminate non-essential information from the entire dataset and deliver the relevant results to the end user. Information filtering is widely used in various domains such as searching Internet, e-mail, e-commerce, multimedia distributed system, blogs, etc. (for a survey, see Hanani et al. \cite{hanani2001information}). There are two basic approaches for IF: content-based filtering and collaborative filtering. Belkin and Croft \cite{belkin1992information} discusses the relationship between IR and IF, and, establish that IF is a special kind of IR with the same set of research challenges and outcomes. They have a common goal to provide relevant information to the user but from different aspects. Hanani et al. \cite{hanani2001information} give a brief overview of IF and discuss the difference between IR and IF with respect to research issues. For improving the relevancy of results obtained after IF, several QE approaches have been published. Relevance feedback techniques expand the user's query in a way that can well reflect the user's interest and needs \cite{allan1996incremental}. Eichstaedt et al. \cite{eichstaedt2002system} combine user's query with system's master query for improving results.  Other techniques include using user's profile \cite{yu2004nonparametric}, using geographical query footprint \cite{fu2005ontology}, using correlated keywords \cite{zimmer2008exploiting}, using links and anchor texts in Wikipedia \cite{arguello2008document}, using text classification in twitter messages \cite{sriram2010short} and using the behavior-patterns of online users \cite{zhang2012behavior,gao2015pattern}. Recently, Wu et al. \cite{wu2016cccf} reformulated the query using the user-item co-clustering method for improving the Collaborative Filtering technique. Another work  by Zervakis et al. \cite{zervakis2017query} reorganizes query using  DBpedia\footnote{http://wiki.dbpedia.org/} and ClueWeb09\footnote{http://lemurproject.org/clueweb09/} corpora for efficient Boolean IF. The proposed approach uses linguistically motivated concepts, such as words, to support continuous queries that are comprised of conjunctions of keywords. These continuous queries may be used as a basis for query languages that support not only basic Boolean operators but also more complex constructs, such as proximity operators and attributes. 
\subsubsection{Multimedia Information Retrieval}
Multimedia Information Retrieval (MIR) deals with searching and extracting the semantic information from multimedia documents, such as audio, video and image (Lew et al. \cite{lew2006content} gives a good survey in this regard). For IR in multimedia documents, most of the MIR systems typically rely on text-based searches, such as title, captions, anchor text, annotations and surrounding HTML or XML depiction. This approach can fail when metadata is absent or when the metadata cannot precisely describe the actual multimedia content. Hence, QE plays a crucial role in extracting the most relevant multimedia data.

Audio retrieval deals with searching audio files in large collections of audio files. The retrieved files should be similar to the audio query, which is in natural language. The search analyzes the actual contents of the audio rather than the metadata such as keywords, tags, and/or descriptions associated with the audio.
For searching spoken audio, a common approach is to do a text search on the transcription of the audio file. However, the transcription is obtained automatically by speech translation software, and hence, contains errors. In such a case, expanding the transcription by adding related words greatly improves the retrieval effectiveness \cite{singhal1999document}. However, for text document retrieval, the benefits of such a document expansion are limited  \cite{wei2007modeling}.
Jourlin et al. \cite{jourlin1999general} show that QE can improve the average precision by 17 percent in audio retrieval. Barrington et al. \cite{barrington2007audio} follow QE technique based on semantic similarity in audio IR. Tejedor et al. \cite{tejedor2012comparison} compare language dependent and language independent queries through examples and conclude that the language dependent setup provides better results in spoken term detection. Recently, Khwileh and Jones \cite{khwileh2016investigating} presented a QE method for social audio contents, where the QE approach uses three speech segments: semantic, window and discourse-based segments.

In video retrieval, queries and documents have both visual as well as textual aspect. The expanded text queries are matched with the manually established text descriptions of the visual concepts. Natsev et al. \cite{natsev2007semantic} expand text query using lexical, statistical and content-based approaches for visual QE. Tellex et al. \cite{tellex2010grounding} expand queries using the corpus of natural language description based on the accurate evaluation of system performance. A more recent work \cite{thomas2016perceptual} uses meta synopsis for video indexing; the meta synopsis contains vital information for retrieving relevant videos. 

In image retrieval, a common approach to retrieve relevant images is querying using textures, shapes, color and visual aspect that match with the image descriptions in the database (for reviews on image retrieval see Li et al. \cite{li2016socializing} and Datta et al. \cite{datta2008image}). Kuo et al. \cite{kuo2009query} present two QE approaches: intra expansion (expanded query is obtained from existing query features) and inter expansion (expanded query is obtained from the search results). Hua et al. \cite{hua2013clickage} use the query logs data for generic web image searches. Borth et al. \cite{borth2013large} use multitag for image retrieval, whereas, Liu et al. \cite{liu2013image} retrieve images using a query adaptive hashing method. Xie et al. \cite{xie2014contextual} present a contextual QE technique to overcome the semantic gap of visual vocabulary quantization, and,  performance and storage loss due to QE in image retrieval. 
\subsubsection{Other Applications}
Some other recent applications of QE are  plagiarism detection \cite{nawab2016ir}, event search \cite{douze2013stable,atefeh2015survey,boer2015knowledge}, text classification \cite{wang2016semantic}, patent retrieval \cite{magdy2011study,mahdabi2014effect,wang2016domain}, dynamic process in IoT \cite{huber2016goal,huber2016using}, classification of e-commerce \cite{jammalamadaka2015query}, biomedical IR \cite{abdulla2016improving}, enterprise search \cite{liu2014exploiting}, code search \cite{nie2016query}, parallel computing in IR \cite{doi:10.1108/EUM0000000007201} and twitter search \cite{kumar2013time,zingla2016short}.

Table \ref{tab:2} summarizes some of the prominent and recent applications of QE in literature based on the above discussion. 
\begin{table}[!h]
	\centering
	\caption{Summary of Research in Applications of Query Expansion \label{tab:2}}{
		
		\begin{tabular}{ | M{2.2cm} | M{3.4cm} | M{3.4cm} | M{3.5cm} | }
			\hline

			\textbf{Research Area} & \textbf{Data Sources} & \textbf{Applications} & \textbf{Publications} \\ \hline 
		Personalized social document & Social annotations, user logs, social tag and bookmarking, social proximity and semantic similarity, word embedding, social context & Enhance document's representation and grant a personalized representation of documents to the user & Zhou et al. 2017 \cite{zhou2017query}, Bouadjenek et al. 2016 \cite{bouadjenek2016persador}, Amer et al. 2016 \cite{amer2016toward}, Mulhem et al. 2016 \cite{Mulhem2016}, Hahm et al. 2014 \cite{hahm2014personalized}, Bouadjenek et al. 2013 \cite{bouadjenek2013laicos}, Zhou et al. 2012 \cite{zhou2012improving}, Bouadjenek et al. 2011 \cite{bouadjenek2011personalized}, Biancalana and Micarelli 2009 \cite{biancalana2009social}  \\ \hline
		Question Answering & FAQs, QA pairs, Social network, WordNet, LOD cloud, community question answering (CQA) archive, query logs and web search &  Respond to user's query with quick concise answers rather than retunring all relevant documents & Molino et al. 2016 \cite{molino2016social}, Cavalin et al. 2016 \cite{cavalin2016building}, Liu et al. 2015 \cite{liu2015complementary}, Shekarpour et al. 2013 \cite{shekarpour2013keyword}, Panovich
		et al. 2012 \cite{panovich2012tie},Bian et al. 2008 \cite{bian2008finding}, Riezler et al. 2007 \cite{riezler2007statistical} \\ \hline
		Cross-Language Information Retrieval & User logs, word embeddings, dictionary translations and co-occurrence terms, linguistic resources & Retrieving information written in a language different from user's query language & Zhou et al. 2016 \cite{zhou2016study}, Bhattacharya et al. 2016 \cite{bhattacharya2016usingword}, Zhou et al. 2015 \cite{zhou2015query}, Gaillard et al. 2010 \cite{gaillard2010query},Cao et al. 2007 \cite{cao2007extending}, Kraaij et al.
		2003 \cite{kraaij2003embedding} \\ \hline
		Information Filtering & User profile, user log, anchor text, Wikipedia, DBpedia corpus, twitter messages  & Searching results on Internet, e-mail, e-commerce and multimedia distributed system & Zervakis et al. 2016 \cite{zervakis2017query}, Wu et al. 2016 \cite{wu2016cccf}, Gao et al. 2015 \cite{gao2015pattern}, Zhang and Zeng 2012 \cite{zhang2012behavior}, Arguello et al. 2008 \cite{arguello2008document}, Fu et al. 2005 \cite{fu2005ontology}, Yu et al. 2004 \cite{yu2004nonparametric}  \\ \hline
		Multimedia Information Retrieval & Title, captions, anchor text, annotations,  meta synopsis, query logs, multitag, corpus of natural language and surrounding html or xml depiction & Searching and extracting semantic information from multimedia documents (audio, video and image) such as audio retrieval, video retrieval and image retrieval & Khwileh and Jones 2016 \cite{khwileh2016investigating}, Thomas et al. 2016 \cite{thomas2016perceptual}, Li et al. 2016 \cite{li2016socializing}, Xie et al. 2014 \cite{xie2014contextual}, Liu et al. 2013 \cite{liu2013image}, Tejedor et al. 2012 \cite{tejedor2012comparison}, Tellex et al. 2010 \cite{tellex2010grounding}, Kuo et al. 2009 \cite{kuo2009query}, Wei and Croft 2007 \cite{wei2007modeling} \\ \hline
		Others & Word embeddings, CLEF- IP patent data, Top documents, Wikipedia, DBpedia, TREC collection, Genomic data sets, top tweets, etc. & Text classification, patent retrieval, plagiarism detection, dynamic process in IoT, twitter search, biomedical IR, code search, event search, enterprise search  & Wang et al. 2016 \cite{wang2016semantic}, Wang and Lin 2016 \cite{wang2016domain}, Nawab et al. 2016 \cite{nawab2016ir}, Huber et al.
		2016 \cite{huber2016goal}, Zingla et al. 2016 \cite{zingla2016short}, Abdulla et al. 2016 \cite{abdulla2016improving}, Nie et al. 2016 \cite{nie2016query}, Atefeh and Khreich 2015 \cite{atefeh2015survey}, Liu et al. 2014 \cite{liu2014exploiting} \\ \hline
	\end{tabular}}
\end{table}

\section{Classification of Query Expansion Approaches}
\label{sec4}
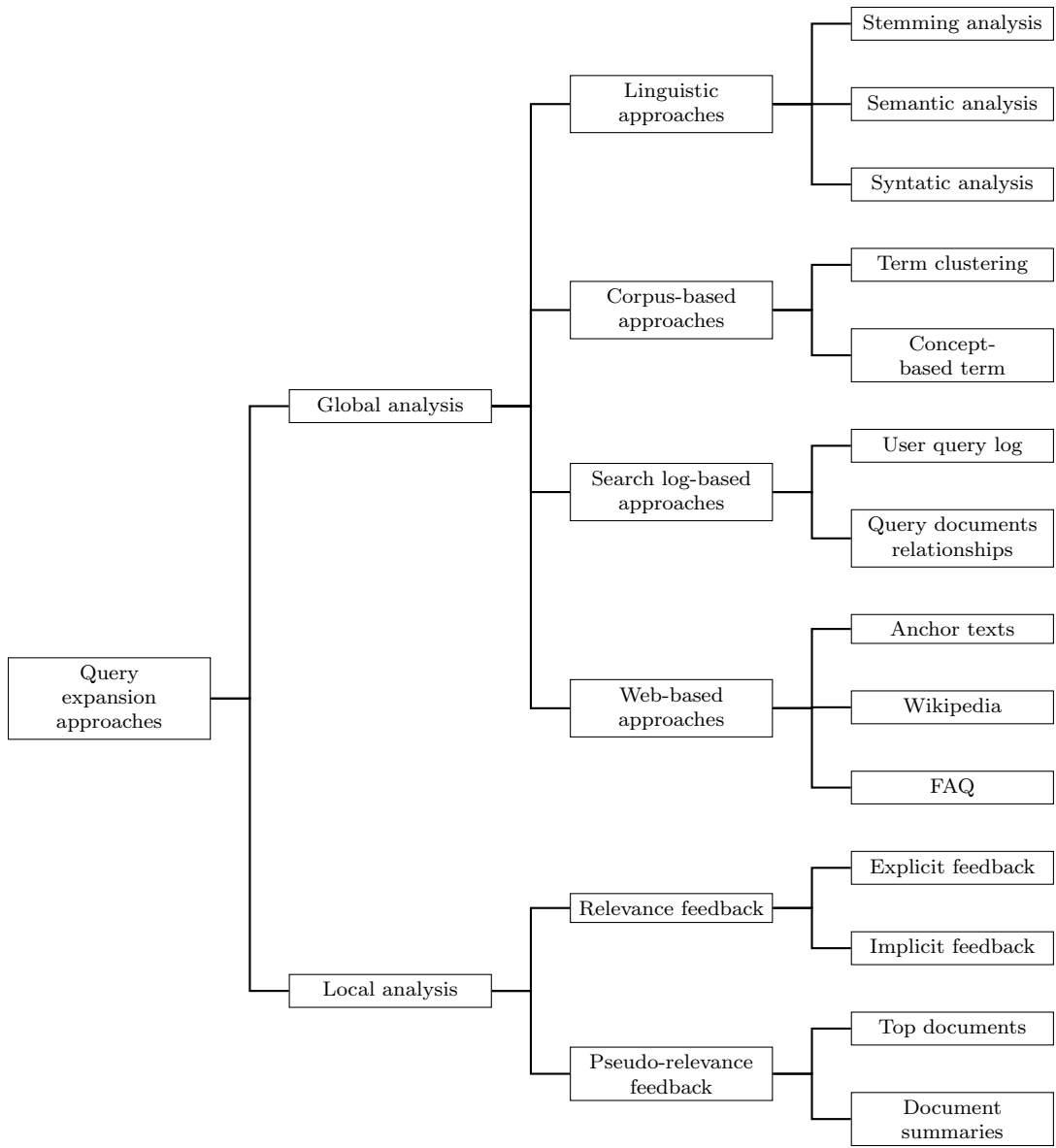
\begin{figure*}     
	\begin{tikzpicture}[grow'=right,level distance=1.50in,sibling distance=.25in]
	\tikzset{edge from parent/.style= 
		{thick, draw, edge from parent fork right},
		every tree node/.style=
		{draw,minimum width=1in,text width=1in,align=center}}
	\Tree 
	[. {Query \\expansion approaches}
	[ .{Global analysis}
	[.{Linguistic approaches}
	[.{Stemming analysis}
	]
	[.{Semantic analysis} ]
	[.{Syntatic analysis} ]
	]
	[.{Corpus-based approaches}
	[.{Term clustering} ]
	[.{Concept-based term} ]
	]
	[.{Search log-based approaches}
	[.{User query log} ]
	[.{Query documents relationships} ]
	]
	[.{Web-based approaches}
	[.{Anchor texts} ]
	[.{Wikipedia} ]
	[.{FAQ} 
	]
	] 
	]
	[.{Local analysis}
	[.{Relevance feedback}
	[.{Explicit feedback} ]
	[.{Implicit feedback} ]
	]
	[.{Pseudo-relevance feedback} 
	[.{Top documents} ]
	[.{Document summaries} ]
	]
	]
	]
	]]
	\end{tikzpicture}
	\caption{Classification of QE approaches based on data sources.}
	\label{fig:6} 
\end{figure*}

On the basis of data sources used in QE, several approaches have been proposed. All these approaches can be classified into two main groups: (1) Global analysis and (2) Local analysis. Global and Local analyses can be further split into four and two subclasses respectively as shown in Fig. \ref{fig:6}. This section discusses the QE approaches based on the properties of various data sources used in QE as shown in  Fig. \ref{fig:6}

\subsection{Global Analysis}
In the global analysis, QE techniques implicitly select expansion terms from hand-built knowledge resources or from large corpora for reformulating the initial query. Only individual query terms are considered for expanding the initial query. The expansion terms are semantically similar to the original terms. Each term is assigned a weight; the expansion terms can be assigned less weight in comparison to the original query terms. Global analysis can be classified into four categories on the basis of query terms and data sources: (i) linguistic-based, (ii) corpus-based, (iii) search log-based, and (iv) web-based. Each approach has been discussed briefly in the following sections.
\subsubsection{Linguistic-based Approaches}
The approaches in this category analyze the expansion features such as lexical, morphological, semantic and syntactic term relationships, to reformulate or expand the initial query terms. They use thesauruses, dictionaries, ontologies, Linked Open Data (LOD) cloud or other similar knowledge resources such as WordNet or ConceptNet. 

Word stemming is one of the first and among the most influential QE approaches in linguistic association to reduce the inflected word from its root word. The stemming algorithm (e.g., \cite{porter1980algorithm}) can be utilized either at retrieval time or at indexing time. When used during retrieval, terms from initially retrieved documents are picked, and then, these terms are harmonized with the morphological types of query terms (e.g., \cite{krovetz1993viewing,paice1994evaluation}). When used during indexing time, document word stems are picked, and then, these words are harmonized with the query root word stems (e.g., \cite{hull1996stemming}). A morphological approach is an ordered way of studying the internal structure of the word. It has been shown to give better results than the stemming approach \cite{bilotti2004works,moreau2007automatic}, however, it requires querying to be done in a structured way.

Use of semantic and contextual analysis are other popular QE approaches in the linguistic association. It includes knowledge sources such as Ontologies, LOD cloud, dictionaries and thesaurus. In the context of ontological based QE, Bhogal et al. \cite{bhogal2007review} use domain-specific and domain-independent ontologies. Wu et al. \cite{wu2011study} utilize the rich semantics of domain ontology and evaluate the trade-off  between the improvement in retrieval effectiveness and the  computational cost. Several research works have been done on QE using a thesaurus. WordNet is a well-known thesaurus for expanding the initial query using word synsets. As discussed earlier, many of the research works use WordNet for expanding the initial query. For example, Voorhees \cite{voorhees1994query} uses WordNet to find the synonyms. Smeaton et al. \cite{smeaton1995trec} use WordNet and POS tagger for expanding the initial query. However, this approach faces some practical problems, such as the absence of accurate matching between query and senses, the absence of proper nouns, and, one query term mapping to many noun synsets and collections. Generally, the utilization of WordNet for QE is beneficial only if the query words are unambiguous in nature \cite{gonzalo1998indexing,voorhees1994query}; using word sense disambiguation (WSD) to remove ambiguity is not easy \cite{navigli2009word,pal2015word}. Several research works have attempted to address the WSD problem. For example, Navigli and Velardi \cite{navigli2005structural} suggest that instead of considering the replacement of the initial query term with its synonyms, hyponyms, and hyperonyms, it is better to extract  similar concepts from the query domain from  WordNet (such as the common nodes and glossy terms). Gong and Cheang \cite{gong2006multi} use the semantically similar information from WordNet present in different groups; this may be combined to expand the initial query. Zhang et al.  \cite{zhang2009concept}, Song et al. \cite{song2007integration}, and Liu et al. \cite{liu2004effective} combine WordNet concepts -- that are extracted by applications of heuristic rules to match similar query terms -- with other term extraction techniques. Shekarpour et al. \cite{shekarpour2013keyword} use linguistic and semantic features of the initial query over linked data for QE as discussed earlier in Sec. \ref{QA}. Recently, Agirre et al. \cite{agirre2014random} introduced a WSD algorithm based on random walks over large Lexical Knowledge Bases (LKB). Their experiments give better results than other graph-based approaches when executed on a graph built from WordNet and eXtended WordNet \cite{mihalcea2001extended}. Nowadays, Word Embeddings techniques are being widely used for QE, e.g., by Roy et al. \cite{roy2016using}, Diaz et al. \cite{diaz2016query} and Kuzi et al. \cite{kuzi2016query} as discussed earlier. 

Another important approach that improves the linguistic information of the initial query is syntactic analysis \cite{zhang2011syntactic}. Syntactic based QE uses the enhanced relational features of the query terms for expanding the initial query. It expands the query mostly through statistical approaches \cite{wu2011study}. It recognizes the term dependency statistically \cite{riezler2007statistical} by employing techniques such as term co-occurrence. Sun et al. \cite{sun2006mining} use this approach for extracting contextual terms and relations from the external corpus. Here, it uses two dependency relation based query
expansion techniques for passage retrieval: Density-based system (DBS) and  Relation based system (RBS). DBS makes use of relation analysis to extract high-quality contextual terms. RBS extracts relation paths for QE in a density and relation-based passage retrieval framework. The syntactic analysis approach may be beneficial for natural language queries in search tasks,  where linguistic analysis can break the task into a sequence of decisions \cite{zhang2011syntactic} or integrate the taxonomic information  effectively \cite{liu2008query}.

\subsubsection{Corpus-based Approaches}
Corpus-based Approaches examine the contents of the whole text corpus to recognize the expansion features to be utilized for QE. They are one of the earliest statistical approaches for QE. They create co-relations between terms based on co-occurrence statistics in the corpus to form sentences, paragraphs, or neighboring words, which are used in the expanded query. Corpus-based approaches have two admissible strategies: (1) term clustering \cite{jones1971automatic,minker1972evaluation,crouch1992experiments}, which groups document terms into clusters based on their co-occurrences, and, (2) concept based terms \cite{qiu1993concept,fonseca2005concept,natsev2007semantic}, where expansion terms are based on the concept of query rather than the original query terms. Kuzi et al. \cite{kuzi2016query} select the expansion terms after the analysis of the corpus using word embeddings, where each term in the corpus is characterized by an embedded vector. Zhang et al. \cite{zhang2016learning} use four corpora as data sources (including one industry and three academic corpora) and present a Two-stage Feature Selection framework (TFS) for QE known as Supervised Query Expansion (SQE) (already discussed in section \ref{sec1}). Some of the other approaches established an association thesaurus based on the whole corpus by using, e.g.,  context vectors\cite{gauch1999corpus}, term co-occurrence\cite{carpineto2001information},  mutual information \cite{hu2006improving} and interlinked Wikipedia articles \cite{milne2008learning}.
\subsubsection{Search log-based Approaches}
These approaches are based on the analysis of search logs. User feedback, which is an important source for suggesting a set of similar terms based on the user's initial query, is generally explored through the analysis of search logs. With the fast growing size of the web and the increasing use  of web search engines, the abundance of search logs and their ease of use  have made them an important source for QE. It usually contains user queries corresponding to the URLs of Web pages. Cui et al. \cite{cui2002probabilistic} use the query logs to extract probabilistic correlations between query terms and document terms. These correlations are further used for expanding the user's initial query. The authors improved upon this in \cite{cui2003query} by using search logs for QE; their experiments give better results when compared with QE based on pseudo-relevance feedback. One of the advantages of using search logs is that it implicitly incorporates relevance feedback. On the other hand, it has been shown by White et al. \cite{white2005study} that implicit measurements are relatively good, however,  their performance may not be the same for all types of users and search tasks.

There are commonly two types of QE approaches used on the basis of web search logs. The first type considers queries as documents and extracts features of these queries that are related to the user's initial query \cite{huang2003relevant}. Among the techniques based on the first approach, some use their combined retrieval results \cite{huang2009analyzing}, while some do not (e.g., \cite{huang2003relevant,yin2009query}).  In the second type of approach, the features are extracted on relational behavior of queries. For example, Bbaeza and Tiberi \cite{baeza2007extracting} represent queries in a graph based vector space model (query-click bipartite graph) and analyze the graph constructed using the query logs. Cui et al. \cite{cui2003query}, Riezler et al. \cite{riezler2007statistical}, and Cao et al. \cite{cao2008context} extract the expansion terms directly from the clicked results. Fitzpatrick and Dent \cite{fitzpatrick1997automatic}, and, Wang and Zhai \cite{wang2007learn} use the top results from the past query terms entered by the users. Under the second approach, queries are also extracted  from related documents \cite{billerbeck2003query,wang2008mining} or through user clicks \cite{xue2004optimizing,yin2009query,hua2013clickage}. The second type of approach is more popular and has been shown to give better results. 

\subsubsection{Web-based Approaches}
These approaches include Wikipedia and anchor texts from websites  for expanding the user's original query. These approaches have gained popularity in recent times.  Anchor text was first used by Mcbryan \cite{mcbryan1994genvl} for associating  hyper-links with linked pages, as well as with the pages in which anchor texts are found. In the context of a web-page, an anchor text can play a role similar to the title since the anchor text pointing to a page can serve as a concise summary of its contents. It has been shown that user search queries and anchor texts are very similar because an anchor text is a brief characterization of its target page. Kraft and Zien \cite{kraft2004mining} used anchor texts for QE; their experimental results suggest that  anchor texts can be used to improve the traditional QE based on query logs. On similar lines, Dang and Croft \cite{dang2010query} suggested that anchor texts can be an effective substitute for query logs. It demonstrated the effectiveness of QE techniques using log-based stemming through experiments on standard TREC collection dataset.

Another popular approach is the use of Wikipedia articles, titles and hyper-links (in-links and out-linsk) \cite{arguello2008document,almasri2013wikipedia}. As we know, Wikipedia is the largest encyclopedia freely available on the web; articles are regularly updated and new ones are added every day. These features make it an ideal knowledge source for QE. Recently, quite a few research works have used it for QE (e.g., \cite{li2007improving,arguello2008document,xu2009query,aggarwal2012query,almasri2013wikipedia}). Al-Shboul and Myaeng \cite{al2014wikipedia} attempt  to enrich initial queries using semantic annotations in Wikipedia articles combined with phrase-disambiguation. Experimental results show better results in comparison to the relevance based language model.

FAQs are another important web-based source of information  for improving the QE. Recently published article by Karan and Snajder \cite{karan2015evaluation} use domain-specific FAQs data for manual QE. Some of the other works using FAQs are  \cite{agichtein2004learning,soricut2006automatic,riezler2007statistical}.

\subsection{Local Analysis}
Local analysis includes QE techniques that select expansion terms from the collection of documents retrieved in response to the user's initial (unmodified) query. The working belief is that the documents retrieved in response to the user's initial query are relevant, hence, terms present in these documents should also be relevant to the initial query. Using local analysis, there are two ways to expand the user's original query: (1) Relevance feedback and (2) Pseudo-relevance feedback. These are discussed next. 
\subsubsection{Relevance Feedback (RF)}
In this approach, the user's feedback about documents retrieved in response to the initial query is collected; the feedback is about  whether or not the retrieved documents are relevant to the user's query. The query is reformulated based on the documents found relevant as per the user's feedback. Rocchio's method \cite{rocchio1971relevance} was amongst the first to use relevance feedback. Relevance feedback can further be categorized into two types: explicit feedback and implicit feedback. In explicit feedback, the user explicitly evaluates the relevance of retrieved documents (as done in \cite{Salton90improvingretrieval,harman1992relevance}), whereas, in implicit feedback, the user's activity on the set of documents retrieved in response to the initial query is used to infer the user's preferences indirectly (e.g., as done in \cite{chirita2007personalized,zhou2012improving,gao2015pattern}). Relevance feedback suffers from  the lack of semantics in the corpus \cite{wu2011study}. This restrains its applications in several occasions, for example, when the query concept is as general as a disjunction of more specific concepts (see Chap. 9 in the book by Manning et al. \cite{Manning:2008:IIR:1394399}). Some of the research works based on relevance feedback are  \cite{buckley1994effect,salton1997improving,ruthven2003survey,Manning:2008:IIR:1394399}; these have been discussed earlier in Sec. \ref{sec:3.2.3}. 
\subsubsection{Pseudo-relevance Feedback (PRF)}
Here, neither explicit nor implicit feedback  of the user is collected. Instead, the feedback collection process is automated by directly using the top-ranked documents (or their snippets) -- retrieved in response to the initial query -- for QE. Pseudo-relevance feedback is also known as  blind feedback, or, retrieval feedback. It has been discussed briefly earlier in Sec. \ref{sec:3.2.3}. This technique was first proposed by Croft and Harper \cite{croft1979using}, who employ this technique in a probabilistic model. Xu and Croft \cite{xu2000improving} proposed ``local context analysis" technique to extract the QE terms from the top documents retrieved in response to the initial query. Each of the candidate expansion terms is assigned a score on the basis of co-occurrence of query terms.  The candidate terms with the highest score are selected for query reformulation. A recent work by Singh and Sharan \cite{singh2016new} uses fuzzy logic-based QE techniques and selects the top-retrieved documents based on PRF. Here, each expansion term is assigned a distinct relevance score using fuzzy rules. Then, the terms having the highest scores are selected for QE. The experimental results demonstrate that the proposed approach achieves significant improvement over individual expansion, expansion on the basis of the entire query and other related advanced methods. ALmasri et al. \cite{almasri2016comparison} proposed deep learning based QE technique and compared it with PRF and other expansion modules; the results show a notable improvement over other techniques using various language models for evaluation.

Considering the top retrieved documents may not always be the best strategy. For example, for a particular query, if the top retrieved documents have very similar contents, the expanded terms -- selected from the top retrieved documents -- will also be very similar. Hence, the expanded terms will not be useful for effective QE. Apart from using the top-ranked documents or their snippets, several other approaches have been proposed. For example, techniques based on passage extraction \cite{xu1996query},  text summarization  \cite{lam2001applying}, and  document summaries \cite{chang2006query}. Some of the other works using PRF are  \cite{cao2008selecting,xu2009query,lv2010positional}; these have been discussed in earlier sections.\\\\
\textbf{Comparative Analysis:} Of all the approaches mentioned earlier, corpus-based approaches are considered more effective than those based on linguistic-based approaches, whether it is global or local analysis. The main reason behind this is that linguistic-based approaches require a concrete linguistic relation (based on sense, meaning, concept etc.) between a query term and a relevant term for the latter to be discovered, while corpus-based approaches can discover the same relevant term simply based on co-occurrences with the query term. Generally, utilization of linguistic-based approaches is beneficial only if the query terms are unambiguous in nature. While several research works have attempted to remove the ambiguity using word sense disambiguation (WSD), exact solutions are very difficult to achieve. For statistical approaches, the local analysis seems to perform better than corpus-based approaches because the extracted features are query specific, whereas techniques based on Web data (such as user query logs or anchor texts) have not yet been systematically evaluated or compared with others on standard test collection.

The use of thesaurus, dictionaries, WordNet or ConceptNet presents some good expansion terms in linguistic-based approaches, but it also causes topic-drift. These approaches can only be used when we know the query's domain or for domain-specific searches because domain-specific resources reduce the topic-drift in such cases. For local analysis, relevance feedback has been demonstrated to be more robust in performance than pseudo-relevance feedback. The primary reason behind this is that pseudo-relevance feedback depends significantly on the execution of the user's initial query; if the initial query is poorly formulated or ambiguous, then the expansion terms extracted from the retrieved documents may not be relevant. Billerbeck et al. \cite{billerbeck2004questioning} reported that pseudo-relevance feedback improved only one-third of queries in their experimental collection. However, based on the recent trends in literature, hybrid techniques (combination of two or more techniques) give best results and seem to be more effective with respect to diversity of users, queries and document corpus. Considering different data sources, an analysis specific to the  local context of a data source can improve retrieval performance by combining global analysis with local feedback. Further, it can be concluded to use the phrase-based expansion technique that uses hybrid data sources for automatic query expansion. In addition to effectiveness and efficiency, we would like to highlight additional pointers for choosing QE technique: (1) Linguistic, Web and Search log-based approaches make use of data that are not always available or suitable for the information retrieval task, (2) Corpus-based approaches are not ideal for dynamic document collection, (3) Query-based approaches are dependent on the quality of the first-pass retrieval documents.

Many studies \cite{koenemann1996case,beaulieu1997experiments,ruthven2003survey} have been done to compare the effectiveness of automatic and interactive query expansions. Intuitively, since the user is the one who decides which document is relevant to his/her query, the user should be able to make a better decision than the system with respect to the terms to be added to the initial query. However, experimental results do not offer conclusive results regarding interactive query expansion being more effective than automatic query expansion. For example, Beaulieu \cite{beaulieu1997experiments} shows that automatic query expansion is more effective than interactive query expansion in the operational setting. On the other hand, Koenemann and Belkin \cite{koenemann1996case} reported user satisfaction and improved system performance with interactive query expansion system. However, Ruthven and Lalmas \cite{ruthven2003survey} did a simulation study and reported that interactive query expansion has potential to achieve higher performance than automatic query expansion, but it is difficult for the users to filter the expansion terms that represent relevant documents.

In summary, there is a wide range of QE approaches that present various characteristics and are mostly useful or applicable in specific circumstances. The best option depends on the evaluation of several factors, including the type of queries, availability and features of external data sources, kind of collection being searched, facilities offered by the underlying weighting and ranking system, and efficiency requirements.

Table \ref{approach} summarizes applicability of QE techniques with respect to the approach and the Data sources used. Applicability and the data sources used by each technique have been presented in a comparative manner; all the techniques have been categorized under two main approaches: Global analysis and Local analysis. 

\begin{table}[!h]
	\centering
	\caption{Applicability of QE techniques categorized with respect to Data sources. \label{approach}}{
		
		\begin{tabular}{ | M{1.6cm} | M{2.0cm} | M{2.2cm} |M{2.7cm}| M{3.9cm} | }
			\hline 
			
			\textbf{Approaches} & \textbf{Sub-Approaches} & \textbf{Data Sources used} & \textbf{Applicability} & \textbf{Publications} \\ \hline 
			\multirow{20}{1.5cm}{\centering {Global Analysis}} & Linguistic approaches & Thesaurus, dictionaries, ontologies, LOD cloud, WordNet, ConceptNet & Word stemming, semantic and contextual analysis, syntactic analysis  & Porter 1980 \cite{porter1980algorithm}, Krovetz 1993 \cite{krovetz1993viewing}, Voorhees 1994 \cite{voorhees1994query}, Bilotti et al. 2004 \cite{bilotti2004works}, Sun et al. 2006 \cite{sun2006mining}, Bhogal et al. 2007 \cite{bhogal2007review}, Zhang et al. 2009 \cite{zhang2009concept}, Wu et al. 2011 \cite{wu2011study}, Agirre et al. 2014 \cite{agirre2014random}, Kuzi et al. 2016 \cite{kuzi2016query}
			\\\cline{2-5}  & Corpus-based approaches & Corpus based thesaurus, text corpus & Term clustering, finding co-relation between terms, mutual information extraction, concept based term extraction & Jones 1971 \cite{jones1971automatic}, Minker et al. 1972 \cite{minker1972evaluation}, Qiu and Frei 1993 \cite{qiu1993concept}, Carpinto et al. 2001 \cite{carpineto2001information}, Fonseca et al. 2005 \cite{fonseca2005concept}, Hu et al. 2006 \cite{hu2006improving}, Natsev et al. 2007 \cite{natsev2007semantic}, Kuzi et al. 2016 \cite{kuzi2016query}
			\\\cline{2-5}  & Search log-based approaches  & Search logs, query logs, user logs  & Features extraction based on relational behavior of user's queries, Query-Documents relationship & Cui et al. 2002 \cite{cui2002probabilistic}, Cui et al. 2003 \cite{cui2003query}, White et al. 2005 \cite{white2005study}, Wang and Zhai 2007 \cite{wang2007learn}, Yin et al. 2009 \cite{yin2009query}, Hua et al. 2013 \cite{hua2013clickage}   \\\cline{2-5}  &Web-based approaches & Wikipedia, anchor texts, FAQs & Enrich initial queries using semantic annotations, Associating hyper-links with linked pages, mutual QE & McBryan 1994 \cite{mcbryan1994genvl}, Kraft and Zien 2004 \cite{kraft2004mining}, Li et al. 2007 \cite{li2007improving}, Xu et al. 2009 \cite{xu2009query}, ALMasri et al. 2013 \cite{almasri2013wikipedia}, Al-Shboul and Myaeng 2014 \cite{al2014wikipedia}, Karan and Snajder 2015 \cite{karan2015evaluation}  \\ \hline
			
			\multirow{7}{1.5cm}{\centering {Local Analysis}} & Relevance feedback & Retrieved documents based upon user's decision & Enrich user's query based on user's feedback & Rocchio  1971 \cite{rocchio1971relevance}, Salton and Buckley 1990 \cite{Salton90improvingretrieval}, Harman 1992 \cite{harman1992relevance}, Salton and Buckley 1997 \cite{salton1997improving}, Chirita et al. 2007 \cite{chirita2007personalized}, Manning et al. 2008 \cite{Manning:2008:IIR:1394399}, Zhou et al. 2012 \cite{zhou2012improving}, Gao et al. 2015 \cite{gao2015pattern}  \\\cline{2-5}  & Pseudo-relevance feedback & Retrieved documents based upon top ranked documents & Enrich user's query based on top ranked documents (instead of user's feedback) retrieved in response to the initial query  & Croft and Harper 1979 \cite{croft1979using}, Xu and Croft 1996 \cite{xu1996query}, Xu and Croft 2000 \cite{xu2000improving}, Lam and Jones 2001 \cite{lam2001applying}, Chang et al. 2006 \cite{chang2006query}, Cao et al. 2008 \cite{cao2008selecting}, Lv and Zhai 2010 \cite{lv2010positional}, ALMasri et al. 2016 \cite{almasri2016comparison}, Singh and Sharan 2016 \cite{singh2016new}
			\\ \hline

	\end{tabular}}
\end{table}

\section{DISCUSSION AND CONCLUSIONS}
\label{sec5}
This article has presented a comprehensive survey highlighting the current progress, emerging research directions, potential new research areas and novel classification of state-of-the-art approaches in the field of QE. The analysis was carried out over four key aspects: (1) data source, which is the collection of documents used for expanding the user's initial query, (2) working methodology, which describes the process for expanding the query, (3) importance and application, which discusses the importance of QE in IR and the use of this technique in the recent trend beyond the key area of IR, and (4) Core approaches, which discuss several QE approaches based on different features of data sources. Furthermore, this article presents a classification of QE approaches into two categories according to the various characteristics of data sources, namely: global analysis, and local analysis. Global analysis was further split into four subcategories: linguistic approaches, corpus-based approaches, search log-based approaches, and Web-based approaches. Local analysis was also split into two subcategories: relevance feedback and pseudo-relevance feedback.

Moreover, the survey provides a discussion of QE in the area of IR as well as the recent trends beyond the IR. QE can be defined as a process in IR that consists of choosing and adding expansion terms to the user's initial query with the goal of minimizing query-document mismatch to improve the retrieval performance. Although there is no perfect solution for the vocabulary mismatch problem in IR systems, QE has the capability to overcome the primary limitations. This is because QE provides the supporting explanation of the information needed for efficient IR, which could not be provided earlier due to the unwillingness or inability of the user. 

As we see in the present scenario of the search systems, most frequent queries are still one, two, or three words; the same as in the past few decades. The lack of query terms increases the ambiguity in choosing among the many possible synonymous meanings of the query terms. This heightens the problem of vocabulary mismatch. This, in turn, has motivated the necessity and opportunity to provide intelligent solutions to the vocabulary mismatch problem. Over the past few decades, a lot of research  has been done in the area of QE based on data sources used, applications, and expansion techniques. This article classifies the various data sources into four categories: documents used in the retrieval process, hand-built knowledge resources, external text collections and resources, and hybrid data sources. Recently, hybrid data sources have been used widely for QE; they are a combination of two or more data sources, more than often, web data being one of them. In research involving web data, Wikipedia is a popular data source because it is freely available and is the largest encyclopedia on the web, where articles are regularly updated, and new articles are added.

Expansion approaches can be manual, automatic or interactive (such as linguistic, corpus-based, web-based, search log-based, RF and PRF); they  expand the user's original query on the basis of  query features and available data sources. Query characteristic depends upon query size, lengths of terms, wordiness, ambiguity, difficulty, and objective; addressing each of these features requires specific approaches. Several experimental studies have also reported a remarkable improvement in  retrieval effectiveness: both with respect to precision and recall. These results are a proof of the advancement of research in QE techniques.

With the ever growing wealth of information available on the Internet, web searching has become an integral part of our lives. Every web user wants personalized information according to their interests and commitments, and hence, IR systems need to personalize search results based on the query and the user's interests. For getting these results, IR systems need a personalized QE approach (QE approach based on personal preference), which learns and uses the user profile to reflect his interests as well as his intent. Such an approach can enhance the retrieval performance. We believe personalization of web search results will play an important part in QE research in future. In personalization of web searches, there are two things that should be taken into account. First, how information is presented or structured on the web, and second, how users interact with different personalized systems. This should affect the way a user-interface is designed so that it allows the system to learn more about the user by collecting information about him. Such collected-information will  play an important role in QE for improving the IR.

Beyond the key area of IR, there are other recent applications where QE techniques are widely used. For example, Personalized Social Documents, Question Answering, Cross-Language Information Retrieval, Information Filtering, Multimedia Information Retrieval, Plagiarism Detection, Enterprise Search, Code Search, Biomedical IR, Classification of E-commerce, and Text Classification.

Finally, it can be said that after decades of research in QE, it has matured greatly. While challenges exist for further research, a lot of real life applications are using state of the art techniques. This article will hopefully help a researcher to better understand QE and its use in the area of IR. 

Tables \ref{tab:8}, \ref{tab:9} and \ref{tab:10} summarize influential query expansion approaches in chronological order on the basis of five prominent features: Data Sources, Term Extraction Methodology, Term representation, Term Selection Methodology, and Weighting Schema. 

\subsection{Epistemological Genesis of this Article}
This section describes how the articles reviewed in this paper were discovered. Survey papers by Bhogal et. al. \cite{bhogal2007review}, Fu et.al. \cite{fu2005ontology}, Manning et. al. \cite{Manning:2008:IIR:1394399}, Carpineto and Romano \cite{carpineto2012survey}, He and Ounis \cite{he2007combining}, Biancalana et. al. \cite{biancalana2013social}, and Paik et. al. \cite{paik2014incremental} were our getting started references. We were aware of these references through our prior exposure to the topic of our survey. Another thing we did while starting our review was to search for the term ``Query Expansion" on Google Scholar. In the search results, by looking at the keywords noted in the relevant papers, we identified further related keywords, such as ``Query Formulation",``Information Retrieval", ``Query Enhancement" and ``Internet Searches".  We searched these keywords on Google scholar looking for prominent papers related to query expansion; paying more attention to the papers published in the last 15-20 years. 

Another common approach we took is: whenever we found an article, say $X$, to be an influential reference in the field, we also went through the papers that have cited $X$ and the papers that have been cited by $X$. We applied this technique to our ``getting started" references, as well as to other prominent papers that we came across during the survey.

We also looked up digital libraries of the journals prominent in the field of ``query expansion"/``information retrieval" such as Knowledge and Information Systems (KAIS), Information Retrieval Journal, ACM Computing Surveys, Information Processing and Management, and  ACM Transactions on Information Systems. We did the same  for prominent conferences such as Text REtrieval Conference (TREC), Knowledge Discovery and Data Mining (KDD), Conference and Labs of the Evaluation Forum (CLEF), Special Interest Group on Information Retrieval (SIGIR) and Forum for Information Retrieval Evaluation (FIRE). 

\begin{landscape}
	\begin{table}
		\centering
		
		\caption{Summary of Research in the area of Query Expansion \label{tab:8}}{
			
			\begin{tabular}{ | M{3cm} | M{3cm} | M{3.2cm} | M{3cm} | M{3cm} | M{3cm} |}
				\hline 
				
				\textbf{Reference} & \textbf{Data Sources} & \textbf{Term Extraction Methodology} & \textbf{Term representation} & \textbf{Term Selection Methodology} & \textbf{Weighting Schema} \\ \hline 
				Robertson 1990 \cite{robertson1990term} & Corpus & All terms in corpus & Individual terms & Swets model & Match function   \\ \hline
				Qiu and Frei 1993 \cite{qiu1993concept} & Corpus & All terms in corpus & Individual terms & Term-Concept similarity & Correlation based weights  \\ \hline
				Voorhees 1994 \cite{voorhees1994query} & WordNet & Synsets \& hyponyms of the query & Individual terms & Hyponym chain length & Vectors multiplication of query and concepts \\ \hline
				Xu and Croft 1996 \cite{xu1996query} & Corpus \& top-ranked documents & Contiguous nouns in top retrieved passages  &  Phrases & Term co-occurrence & Ranked-based weights \\ \hline
				Robertson et al. 1999 \cite{robertson1999okapi} & Top-ranked documents  & All terms in top retrieved documents & Individual terms & Robertson selection value (RSV) & Probabilistic reweighting \\ \hline
				Carpineto et al. 2001 \cite{carpineto2001information} & Corpus \& top-ranked documents & All terms in top retrieved documents  & Individual terms & Kullback-Leibler divergence (KLD) & Rocchio \& KLD scores \\ \hline
				Zhai and Lafferty 2001 \cite{zhai2001model} & Corpus \& top-ranked documents & All terms in top retrieved documents  & Individual terms & Mixture model & Query language model \\ \hline
				Lavrenko and Croft 2001 \cite{lavrenko2001relevance} & Corpus \& top-ranked documents  & All terms in top retrieved documents  & Individual terms & Relevance model & Query language model \\ \hline
				Cui et al. 2003 \cite{cui2003query} & User logs \& corpus & Query-documents correlation  & Individual terms & Probabilistic term-term association & Cohesion weights \\ \hline
				Billerbeck et al. 2003 \cite{billerbeck2003query} & Query logs & Query association   & Individual terms & Robertson selection value (RSV) & Probabilistic reweighting \\ \hline
				Kraft and Zien 2004 \cite{kraft2004mining} & Anchor texts & Adjacent terms in anchor text  & Phrases & Median rank aggregation & Unweighted terms \\ \hline
				Liu et al. 2004 \cite{liu2004effective} & Corpus, top-ranked documents \& WordNet & Phrase classification \& WordNet concepts  & Individual terms \& phrases & Term co-occurrence \& WSD & Boolean query \\ \hline
				Bai et al. 2005 \cite{bai2005query} & Top-ranked documents  & Adjacent terms in top-ranked documents  & Individual terms & Term co-occurrence \& Information Flow (IF)  & Query language model \\ \hline
				Collins  and Callan 2005 \cite{collins2005query} & WordNet, corpus, stemmer and top-ranked documents  & Probabilistic term association network & Individual terms & Markov chain & Structured query \\ \hline
				Sun et al. 2006 \cite{sun2006mining} & Corpus  & Relevant contextual terms  & Phrases  & DBS \& RBS & Correlation-based weights \\ \hline
				Hsu et al. 2006 \cite{hsu2006query} & ConceptNet \& WordNet  & Terms having the same concept  & Individual terms  & Using discrimination ability \& concept diversity & Correlation-based weights \\ \hline
				Riezler et al. 2007 \cite{riezler2007statistical} & FAQ data  & Phrases in FAQ answers  & Phrases & SMT techniques & Unweighted terms \\ \hline
				Metzler and Croft 2007 \cite{metzler2007latent} & Corpus \& top-ranked documents & Markov random fields model  & Individual terms & Maximum likelihood & Expanded query graph \\ \hline
				Bai et al. 2007 \cite{bai2007using} & Corpus, user domains \& top-ranked documents & Terms \& nearby terms  & Individual terms & Query classification \& mutual information & Query language model \\ \hline
			\end{tabular}
		}
	\end{table}
\end{landscape}

\begin{landscape}
	\begin{table}
		\centering
		
		\caption{Summary of Research in the area of QE (Cont. from Table \ref{tab:8}) \label{tab:9}}{
			
			\begin{tabular}{ | M{3cm} | M{3cm} | M{3.2cm} | M{3cm} | M{3cm} | M{3cm} |}
				\hline 
				
				\textbf{Reference} & \textbf{Data Sources} & \textbf{Term Extraction Methodology} & \textbf{Term representation} & \textbf{Term Selection Methodology} & \textbf{Weighting Schema} \\ \hline 
				Lee et al. 2008 \cite{lee2008cluster} & Corpus \& top-ranked documents & Clustering of top-ranked documents & Individual terms & Relevance model &  Query language model  \\ \hline
				Cao et al. 2008 \cite{cao2008selecting} & Corpus \& top-ranked documents & All terms in top retrieved documents & Individual terms & Term classification & Query language model  \\ \hline
				Arguello et al. 2008 \cite{arguello2008document} & Wikipedia & Anchor texts in top retrieved Wikipedia documents  & Phrases & Document rank \& link frequency & Sum of entry likelihoods \\ \hline
				Xu et al. 2009 \cite{xu2009query} & Wikipedia & All terms in top retrieved articles  &  Individual terms & Relevance model & Query language model\\ \hline
				Yin et al. 2009 \cite{yin2009query} & Query logs \& Snippets  & All terms in top retrieved snippets \& random walks on query-URL graph & Individual terms & Relevance model \& mixture model  & Query language model \\ \hline
				Dang and Croft 2010 \cite{dang2010query} & Anchor texts & Adjacent terms in anchor text  & Individual terms \& Phrase & Kullback-Leibler divergence (KLD) & Substitution probability \\ \hline
				Lv and Zhai	2010 \cite{lv2010positional} & Corpus \& top-ranked documents & All terms in top retrieved feedback documents  & Individual terms & Positional relevance model (PRM) & Probabilistic reweighting \\ \hline
				Kim et al. 2011 \cite{kim2011automatic} & Corpus  & All terms in top retrieved documents  & Individual terms & Decision tree-based & Boolean query \\ \hline
				Bhatia et al. 2011 \cite{bhatia2011query} & Corpus & All terms in the corpus  & Individual terms \& Phrases & Document-centric approach & Correlation based weights \\ \hline
				Miao et al. 2012 \cite{miao2012proximity} & Corpus & All Proximity-based terms in the corpus   & Individual terms & Proximity-based feedback model (PRoc) & KLD scores \\ \hline
				Zhou et al. 2012 \cite{zhou2012improving} & User logs & Associated terms extracted from top ranked documents  & Individual terms & Annotations and resources the user has bookmarked & Correlation based weights \\ \hline
			    Aggarwal and Buitelaar 2012	\cite{aggarwal2012query} & Wikipedia \& DBpedia. & Top ranked articles from best selected articles as concept candidates  & Individual terms \& phrases & Explicit Semantic Analysis (ESA) score & tf-idf  \\ \hline
				ALMasri et al. 2013 \cite{almasri2013wikipedia} & Wikipedia  & In-link \& out-link articles in top retrieved articles & Individual terms & Semantic similarity & Semantic similarity score \\ \hline
				Pal et al. 2013 \cite{pal2013query} & Corpus  & All terms in top retrieved documents  & Individual terms  & Association \& distribution based term selection & KLD score \\ \hline
				Bouchoucha et al. 2013 \cite{bouchoucha2013diversified} & ConceptNet  & Diverse expansion terms from retrieved documents  & Individual terms &  MMRE (Maximal Marginal Relevance -based Expansion) & MMRE score \\ \hline
				Augenstein et al. 2013 \cite{augenstein2013mapping} & LOD cloud & Neighbors term in whole graph of LOD   &  Individual terms & Mapping Keywords & tf-idf\\ \hline

			\end{tabular}
		}
	\end{table}
\end{landscape}

\begin{landscape}
	\begin{table}
		\centering
		
		\caption{Summary of Research in the area of QE (Cont. from Table \ref{tab:9}) \label{tab:10}}{
			
			\begin{tabular}{ | M{3cm} | M{3cm} | M{3.2cm} | M{3cm} | M{3cm} | M{3cm} |}
				\hline 
				
				\textbf{Reference} & \textbf{Data Sources} & \textbf{Term Extraction Methodology} & \textbf{Term representation} & \textbf{Term Selection Methodology} & \textbf{Weighting Schema} \\ \hline 
		        Pal et al. 2014	\cite{pal2014improving} & WordNet & Synonyms, holonyms \& meronyms of the query terms & Individual terms & Pseudo relevant documents & Normalized weights  \\ \hline
				Paik et al. 2014 \cite{paik2014incremental} & Top retrieved set of documents & All terms in feedback documents & Individual terms & Incremental Blind Feedback (IBF) & Scoring function  \\ \hline
				Al-Shboul and Myaeng 2014 \cite{al2014wikipedia} & Wikipedia & Titles \& in/out links of Wikipedia pages  & Individual terms \& Phrases & Wikipedia page surrogates & Phrase likelihood model\\ \hline
				Dalton et al. 2014 \cite{dalton2014entity} & Wikipedia \& Freebase & All terms in  top ranked articles links to knowledge bases  & Individual terms \& phrases  & Entity query feature expansion (EQFE) & Document retrieval probability \\ \hline
			    Anand and Kotov 2015 \cite{anand2015empirical} & DBpedia \& ConceptNet  & Top-k terms in term association graphs  & Individual terms & Information theoretic measures
				based on co-occurrence \& mixture model  & KLD score \\ \hline
				Xiong and Callan 2015 \cite{xiong2015query} & Freebase & All terms in top retrieved documents  & Individual terms  & tf.idf based PRF \& Freebase's entity
				categories & tf.idf \& Jensen-Shannon divergence score \\ \hline
				Gan and Hon	2015 \cite{gan2015improving} & Wikipedia \& corpus  & Top-k correlated terms in the corpus  & Individual terms &  Markov network model &  Probabilistic query reweighting \\ \hline
				Roy et al. 2016 \cite{roy2016using} & Corpus  & All nearby terms at word
				embedding framework  & Individual terms & K-nearest neighbor approach (KNN) & Query language model \\ \hline
				Diaz et al. 2016 \cite{diaz2016query} & Corpus  & All terms in locally-trained
				word embeddings  & Individual terms & local embedding approach & Query language models \& KLD score \\ \hline
				Singh and Sharan 2016 \cite{singh2016new} & Top-ranked documents & All the unique terms of top N retrieved documents  & Individual terms  & Fuzzy-logic-based approach & CHI score, Co-occurrence score, KLD score \& RSV score \\ \hline
				Zhang et al. 2016 \cite{zhang2016learning} & Corpora  & Top-k terms in retrieved documents  & Individual terms  & Two-stage Feature Selection (TFS) & Probabilistic query reweighting \\ \hline
		    	Zhou et al. 2017 \cite{zhou2017query} & User profiles \& folksonomy data & All semantically similar terms inside the enriched user profiles   & Individual terms & Word embeddings \& topic models & Topical weighting scheme \\ \hline

			\end{tabular}
		}
	\end{table}
\end{landscape}

\begin{acknowledgements}
This publication is an outcome of the R\&D work undertaken in the project under the Visvesvaraya PhD Scheme of Ministry of Electronics \& Information Technology, Government of India, being implemented by Digital India Corporation (formerly Media Lab Asia).
\end{acknowledgements}

\bibliographystyle{spmpsci}      
\bibliography{springer-bibfile}

%
%


\end{document}